# INFERENCE FOR THE LIMITING CLUSTER SIZE DISTRIBUTION OF EXTREME VALUES

By Christian Y. Robert

*CNAM and CREST, France*

Any limiting point process for the time normalized exceedances of high levels by a stationary sequence is necessarily compound Poisson under appropriate long range dependence conditions. Typically exceedances appear in clusters. The underlying Poisson points represent the cluster positions and the multiplicities correspond to the cluster sizes. In the present paper we introduce estimators of the limiting cluster size probabilities, which are constructed through a recursive algorithm. We derive estimators of the extremal index which plays a key role in determining the intensity of cluster positions. We study the asymptotic properties of the estimators and investigate their finite sample behavior on simulated data.

**1. Introduction.** Many results in extreme value theory may be naturally discussed in terms of point processes. Typically, the distribution of extreme order statistics may be obtained by considering the point process of exceedances of a high level. More formally, let $(X_n)$ be a strictly stationary sequence of random variables (r.v.s) with marginal distribution $F$. We assume that for each $\tau > 0$ there exists a sequence of levels $(u_n(\tau))$ such that $\lim_{n\to\infty} n\bar{F}(u_n(\tau)) = \tau$, where $\bar{F} = 1 - F$. It is necessary and sufficient for the existence of such a sequence that $\lim_{x\to x_f} \bar{F}(x)/\bar{F}(x-) = 1$, where $x_f = \sup\{u : F(u) < 1\}$ (see Theorem 1.7.13 in [28]). A natural choice is given by $u_n(\tau) = F^{\leftarrow}(1 - \tau/n)$, where $F^{\leftarrow}$ is the generalized inverse of $F$, that is, $F^{\leftarrow}(y) = \inf\{x \in R : F(x) \geq y\}$. The point process of time normalized exceedances $N_n^{(\tau)}(\cdot)$ is defined by $N_n^{(\tau)}(B) = \sum_{i=1}^n 1_{\{i/n \in B, X_i > u_n(\tau)\}}$ for any Borel set $B \subset E := (0, 1]$. The event that $X_{n-k+1:n}$, the $k$th largest of $X_1, \ldots, X_n$, does not exceed $u_n(\tau)$ is equivalent to $\{N_n^{(\tau)}(E) < k\}$ and the asymptotic distribution of $X_{n-k+1:n}$ is easily derived from the asymptotic distribution of $N_n^{(\tau)}(E)$.

If $(X_n)$ is a sequence of independent and identically distributed (i.i.d.) r.v.s, $N_n^{(\tau)}$ converges in distribution to a homogeneous Poisson process with intensity $\tau$ (see, e.g., [13], Theorem 5.3.2). If the i.i.d. assumption is relaxed and a long range dependence condition is assumed [$\Delta(u_n(\tau))$ defined below], the limiting point process is necessarily a homogeneous compound Poisson process with intensity $\theta\tau$ ($\theta \geq 0$) and limiting cluster size distribution $\pi$ [24]. The constant $\theta$ is referred to as the extremal index and its reciprocal is equal to the mean of $\pi$ under some mild additional assumptions (see [36, 38] for some counterexamples). It may be shown that $\theta \leq 1$ and that the compound Poisson limit becomes Poisson when $\theta = 1$.

If $\lim_{n\to\infty} P(N_n^{(\tau)}(E) = 0) = e^{-\theta\tau}$, then a necessary and sufficient condition for convergence of $N_n^{(\tau)}$ is convergence of the conditional distribution of $N_n^{(\tau)}(B_n)$ with $B_n = (0, q_n/n]$ given that there is at least one exceedance of $u_n(\tau)$ in $\{1, \ldots, q_n\}$ to $\pi$, that is,

$$(1.1) \qquad \lim_{n\to\infty} P(N_n^{(\tau)}(B_n) = m | N_n^{(\tau)}(B_n) > 0) = \pi(m), \qquad m \geq 1,$$

where $(q_n)$ is a $\Delta(u_n(\tau))$-separating sequence (see Section 3). Moreover, if the long range dependence condition $\Delta(u_n(\tau))$ holds for each $\tau > 0$, then $\theta$ and $\pi$ do not depend on $\tau$.

The natural approach to do inference on $\theta$ and $\pi$ is to identify the clusters of exceedances above a high threshold, then to evaluate for each cluster the characteristic of interest and to construct estimates from these values. The two common methods that are used to define clusters are the blocks and runs declustering schemes. The blocks declustering scheme consists in choosing a block length $r_n$ and partitioning the $n$ observations into $k_n = \lfloor n/r_n \rfloor$ blocks, where $\lfloor x \rfloor$ denotes the integer part of $x$. Each block that contains an exceedance is treated as one cluster. The runs declustering scheme consists in choosing a run length $p_n$, and stipulating that any pair of extreme observations separated by fewer than $p_n$ nonextreme observations belong to the same cluster. The block length $r_n$ and the run length $p_n$ are termed the cluster identification scheme sequences and play a key role in determining the asymptotic properties of the estimators.

The problem of inference on the extremal index has received a lot of attention in the literature. The first blocks and runs estimators were constructed by using different probabilistic characterizations of the extremal index (see [13], Section 8.1, [1, 39]). They are determined by two sequences: the sequence of the thresholds $u_n(\tau)$ and the cluster identification scheme sequence. Their major drawback is their dependence on the threshold which is based on the unknown stationary distribution. Estimating this threshold is intricate since, by definition, it is exceeded by very few observations [12]. To circumvent this issue, lower thresholds have to be considered. The following characterizations (see [27, 31])

$$\theta = \lim_{n\to\infty} s_n P\Big(\max_{1\leq i\leq r_n} X_i > u_{s_n}(\tau)\Big)\Big/(r_n\tau),$$

and

$$\theta = \lim_{n\to\infty} P\Big(\max_{2\leq i\leq p_n} X_i \leq u_{s_n}(\tau)\Big|X_1 > u_{s_n}(\tau)\Big),$$

where $s_n = o(n)$, $r_n = o(s_n)$ and $p_n = o(s_n)$, have motivated other blocks and runs estimators [21, 22, 43]; the threshold $u_{s_n}(\tau)$ can be estimated by $X_{n-\lfloor n\tau/s_n\rfloor:n}$. Note that the estimators are determined by two sequences as well: $r_n$ (or $p_n$) and $s_n$. More recently, new methods for identifying clusters of extreme values have been introduced in [26] and new estimators of the extremal index which are less sensitive to cluster identification scheme sequences have been derived. However, to exploit these methods, it is necessary to know whether the process exhibits either an autoregressive or volatility driven dependence structure and to choose an additional threshold to identify the clusters. In order to eliminate the cluster identification scheme sequences, [16] (see also [15]) proposes estimators which are based on the sequence of the thresholds $u_{r_n}(\tau)$ and on inter-exceedance times: a least-squares estimator, a maximum-likelihood estimator and a moment estimator. It is established that the last-mentioned estimator is weakly consistent for $m$-dependent stationary sequences.

There are very few papers which investigate the inference for the limiting cluster size probabilities. In [21], condition (1.1) is used to motivate the following blocks estimators

$$(1.2) \qquad \hat{\pi}_{n,1}(m;r_n,u_{s_n}(\tau)) = \frac{\sum_{j=1}^{k_n} 1_{\{Y_{n,j}(u_{s_n}(\tau))=m\}}}{\sum_{j=1}^{k_n} 1_{\{Y_{n,j}(u_{s_n}(\tau))>0\}}},$$

where $Y_{n,j}(u_{s_n}(\tau)) = \sum_{i=(j-1)r_n+1}^{jr_n} 1_{\{X_i>u_{s_n}(\tau)\}}$, $s_n = o(n)$ and $r_n = o(s_n)$. Let $E_\pi(T) = \sum_{m=1}^{\infty} T(m)\pi(m)$, where $T$ is a function supported on $\{1,2,\ldots\}$. The weak consistency of the estimators

$$\sum_{m=1}^{r_n} T(m)\hat{\pi}_{n,1}(m;r_n,X_{n-\lfloor n\tau/s_n\rfloor:n})$$

of $E_\pi(T)$ is established. Note that they are determined by two sequences: $r_n$ and $s_n$. In [23] the following quantities are considered

$$\hat{\pi}_{n,2}(m;r_n,u_{s_n}(\tau)) = \frac{\sum_{j=1}^{k_n}(R_j^{(m)} - R_j^{(m+1)})1_{\{Y_{n,j}(u_{s_n}(\tau))>0\}}}{\sum_{j=1}^{k_n} 1_{\{Y_{n,j}(u_{s_n}(\tau))>0\}}},$$

where $R_j^{(m)} = \bar{F}(M_j^{(1)})/\bar{F}(M_j^{(m)})$ and $M_j^{(m)}$ is the $m$th largest value of $X_i$, $i = (j-1)r_n + 1, \ldots, jr_n$. A partial comparison with $\hat{\pi}_{n,1}(m; r_n, u_{s_n}(\tau))$ is made under the assumption that $F$ is known. Recently a new method has been proposed in [15]: a recursive algorithm forms estimates of the limiting cluster size probabilities from empirical moments which are based on the joint distributions of the inter-exceedance times separated by other inter-exceedance times. These estimators are only determined by selecting the sequence of thresholds $u_{r_n}(\tau)$. A consistency result for $m$-dependent stationary sequences is given.

In the present paper we introduce new blocks estimators of the limiting cluster size probabilities. The approach is the following. First we estimate the compound probabilities of the limiting point process. Second we use a declustering (decompounding) algorithm to form estimates of the limiting cluster size probabilities. This idea has been proposed recently in [5] and [6] where it is assumed that a sample of the compound Poisson distribution is observed (which is unfortunately not the case here).

More specifically, let us denote by $N_E^{(\tau)}$ the weak limit of $N_n^{(\tau)}(E)$ as $n \to \infty$ when it exists and by $p^{(\tau)} = (p^{(\tau)}(m))_{m \geq 0}$ its distribution. Let $(\zeta_i)_{i \geq 1}$ be a sequence of positive i.i.d. integer-valued r.v.s with distribution $\pi$ and $\eta(\theta\tau)$ be a r.v. with Poisson distribution and parameter $\theta\tau$ such that $\eta(\theta\tau)$ is independent of the $(\zeta_i)_{i \geq 1}$. We have $N_E^{(\tau)} \stackrel{d}{=} \sum_{i=1}^{\eta(\theta\tau)} \zeta_i$, with the convention that the sum equals 0 if the upper index is smaller than the lower index. The distribution of $N_E^{(\tau)}$ is given by

(1.3) $\quad p^{(\tau)}(0) = P(\eta(\theta\tau) = 0) = e^{-\theta\tau},$

(1.4) $\quad p^{(\tau)}(m) = \sum_{j=1}^{m} P(\eta(\theta\tau) = j) P\left(\sum_{i=1}^{j} \zeta_i = m\right) = \sum_{j=1}^{m} \frac{e^{-\theta\tau}(\theta\tau)^j}{j!} \pi^{*j}(m),$

$m \geq 1$, where $\pi^{*j}$ is the $j$th convolution of $\pi$, that is,

$$\pi^{*j}(m) = \begin{cases} 0, & m < j, \\ \sum_{i_1 + \cdots + i_j = m} \pi(i_1) \cdots \pi(i_j), & m \geq j. \end{cases}$$

In risk theory the aggregate claim amount is often assumed to have a compound Poisson distribution. Panjer's algorithm [32] is a method to compute recursively the aggregate claims distribution when the distribution of a single claim is discrete and the distribution of the number of claims is Poisson, Binomial or Negative-Binomial. For the limiting compound Poisson distribution (1.3)–(1.4), the recursion is given by

$$p^{(\tau)}(0) = e^{-\theta\tau},$$

$$p^{(\tau)}(m) = -\frac{\ln(p^{(\tau)}(0))}{m} \sum_{j=1}^{m} j\pi(j) p^{(\tau)}(m-j), \qquad m \geq 1.$$

Note that the $p^{(\tau)}(m)$ can be expressed as a function of the $\pi(j)$, $j = 1, \ldots, m$. It is possible to reverse the algorithm and to evaluate recursively the $\pi(m)$ from the $p^{(\tau)}(j)$, $j = 0, \ldots, m$, and the $\pi(j)$, $j = 0, \ldots, m-1$, in the following way

(1.5)
$$\pi(m) = -\frac{(p^{(\tau)}(m) + m^{-1} \ln(p^{(\tau)}(0)) \sum_{j=1}^{m-1} j\pi(j) p^{(\tau)}(m-j))}{\ln(p^{(\tau)}(0)) p^{(\tau)}(0)},$$

$$m \geq 1.$$

Hence, the inversion of Panjer's algorithm provides an appealing recursive method to estimate the limiting cluster size probabilities.

The content of the paper is organized as follows. In Section 2 we explain how we construct the estimators of the limiting cluster size probabilities. We also derive estimators of the extremal index. We emphasize that all our estimators are determined by one sequence and one (or two) parameter(s). In Section 3 we present and discuss technical conditions which are required for establishing the asymptotic properties. In Section 4 we give results on weak convergence of the estimators. In Section 5 we investigate the finite sample behavior of the estimators on simulated data and we make a comparison with existing estimators. Proofs are gathered in a last section.

**2. Defining the estimators.** In the remainder of the paper we assume that $u_n(\tau) = F^{\leftarrow}(1 - \tau/n)$. The present approach to estimating the limiting cluster size distribution is based on the blocks declustering scheme. We divide $\{1, \ldots, n\}$ into $k_n$ blocks of length $r_n$, $I_j = \{(j-1)r_n + 1, \ldots, jr_n\}$ for $j = 1, \ldots, k_n$, and a last block $I_{k_n+1} = \{r_n k_n + 1, \ldots, n\}$. The number of observations above the threshold $u_{r_n}(\tau)$ within the $j$th block is denoted by

$$N_{r_n,j}^{(\tau)} = \sum_{i \in I_j} 1_{\{X_i > u_{r_n}(\tau)\}}, \qquad j = 1, \ldots, k_n.$$

Since $\lim_{n \to \infty} E(N_{r_n,j}^{(\tau)}) = \tau$, the parameter $\tau$ can be interpreted as the asymptotic mean number of observations which exceed the level $u_{r_n}(\tau)$ for each block. The empirical distribution, $p_n^{(\tau)}$, of the number of exceedances within a block is given by

$$p_n^{(\tau)}(m) = \frac{1}{k_n} \sum_{j=1}^{k_n} 1_{\{N_{r_n,j}^{(\tau)} = m\}}, \qquad m \geq 0.$$

As mentioned in the introduction, the main issue when using these quantities for estimating $p^{(\tau)}$ is that the threshold $u_{r_n}(\tau)$ is based on the unknown stationary distribution. It has to be estimated from the data. We define the estimator of $p^{(\tau)}$ by

$$\hat{p}_n^{(\tau)}(m) = \frac{1}{k_n} \sum_{j=1}^{k_n} 1_{\{\hat{N}_{r_n,j}^{(\tau)} = m\}}, \qquad m \geq 0,$$

where $\hat{N}_{r_n,j}^{(\tau)} = \sum_{i \in I_j} 1_{\{X_i > \hat{u}_{r_n}(\tau)\}}$ and $\hat{u}_{r_n}(\tau) = X_{k_n r_n - \lfloor k_n \tau \rfloor : k_n r_n}$.

Let us now consider the estimators of the limiting cluster size probabilities. To ensure that the entries in (1.5) are nonnegative and that their sum does not exceed 1, we define recursively

$$\hat{\pi}_n^{(\tau)}(m) = \max\left(0, \min\left(\chi_n^{(\tau)}(m), 1 - \sum_{j=1}^{m-1} \hat{\pi}_n^{(\tau)}(j)\right)\right), \qquad m \geq 1,$$

where

$$\chi_n^{(\tau)}(m) = -\frac{(\hat{p}_n^{(\tau)}(m) + m^{-1} \ln(\hat{p}_n^{(\tau)}(0)) \sum_{j=1}^{m-1} j \hat{\pi}_n^{(\tau)}(j) \hat{p}_n^{(\tau)}(m-j))}{\ln(\hat{p}_n^{(\tau)}(0)) \hat{p}_n^{(\tau)}(0)}.$$

We also define smoothed versions by

$$\hat{\bar{\pi}}_n(m) = \frac{1}{\phi - \sigma} \int_\sigma^\phi \hat{\pi}_n^{(\tau)}(m) \, d\tau, \qquad m \geq 1,$$

for given $0 < \sigma < \phi$ (see [35] for a similar averaging technique used to reduce the asymptotic variance of the moment estimator of the extreme value parameter).

Finally, let us derive estimators of the extremal index. This parameter appears in different moments of the distributions of $N_E^{(\tau)}$ and $\zeta_1$ (when they exist)

$$P(N_E^{(\tau)} = 0) = e^{-\theta \tau}, \qquad E(\zeta_1) = \theta^{-1}, \qquad V(N_E^{(\tau)}) = \theta \tau E(\zeta_1)^2.$$

Fix an integer $m \geq 1$. We consider two approximations of $\theta$

$$\theta_2(m) = \frac{1}{\sum_{j=1}^m j \pi(j)} \quad \text{and} \quad \theta_3^{(\tau)}(m) = \frac{\sum_{j=0}^m (j-\tau)^2 p(j)}{\tau \sum_{j=1}^m j^2 \pi(j)}.$$

Estimators of $\theta$, $\theta_2(m)$ and $\theta_3^{(\tau)}(m)$ can be constructed by equating theoretical moments to their empirical counterparts

$$\hat{\theta}_{1,n}^{(\tau)} = -\frac{\ln(\hat{p}_n^{(\tau)}(0))}{\tau}, \qquad \hat{\theta}_{2,n}^{(\tau)}(m) = \frac{1}{\sum_{j=1}^m j \hat{\pi}_n^{(\tau)}(j)},$$

$$\hat{\theta}_{3,n}^{(\tau)}(m) = \frac{\sum_{j=0}^m (j-\tau)^2 \hat{p}_n^{(\tau)}(j)}{\tau \sum_{j=1}^m j^2 \hat{\pi}_n^{(\tau)}(j)}.$$

$\hat{\theta}_{1,n}^{(\tau)}$ can be seen as a slight modification of the estimator in equation (1.5) in [39]. $\hat{\theta}_{2,n}^{(\tau)}(m)$ has been studied in [21] with (1.2) as an estimator of the limiting cluster size distribution and $m = r_n$. To the best of our knowledge, $\hat{\theta}_{3,n}^{(\tau)}(m)$ seems to be new. Finally, let us define $\widehat{\bar{\theta}}_{1,n}$ by the smoothed version of the first estimator

$$\widehat{\bar{\theta}}_{1,n} = \frac{1}{\phi - \sigma} \int_\sigma^\phi \hat{\theta}_{1,n}^{(\tau)} d\tau.$$

All estimators (resp. smoothed versions of the estimators) introduced in this section are determined by the sequence $r_n$ and the parameter $\tau$ (resp. $\phi$ and $\sigma$). They provide an interesting alternative to the estimators introduced in [16] and [15] where it is only needed to select the sequence of the thresholds $u_{r_n}(\tau)$. Note that both methods share the same parsimony since in our case $u_{r_n}(\tau)$ is estimated by $X_{k_n r_n - \lfloor k_n \tau \rfloor : k_n r_n}$.

**3. Technical conditions.** In this section we present and discuss technical conditions which are required for establishing the asymptotic properties of the estimators. We begin by giving definitions which are essentially due to [20, 27, 33].

The stationary sequence $(X_n)$ is said to have extremal index $\theta \geq 0$ if, for each $\tau > 0$, $\lim_{n \to \infty} P(N_n^{(\tau)} = 0) = \exp(-\theta\tau)$.

Fix an integer $r \geq 1$ and $\tau_1 > \cdots > \tau_r > 0$. Define $\mathcal{F}_{p,q}^{(\tau_1,\ldots,\tau_r)}$ as the $\sigma$-algebra generated by the events $\{X_i > u_n(\tau_j)\}$, $p \leq i \leq q$ and $1 \leq j \leq r$, and write

$$\alpha_{n,l}(\tau_1,\ldots,\tau_r) \equiv \sup\{|P(A \cap B) - P(A)P(B)| :$$
$$A \in \mathcal{F}_{1,t}^{(\tau_1,\ldots,\tau_r)}, B \in \mathcal{F}_{t+l,n}^{(\tau_1,\ldots,\tau_r)}, 1 \leq t \leq n - l\}.$$

The condition $\Delta(\{u_n(\tau_j)\}_{1 \leq j \leq r})$ is said to hold if $\lim_{n \to \infty} \alpha_{n,l_n}(\tau_1,\ldots,\tau_r) = 0$ for some sequence $l_n = o(n)$. The long range dependence condition $\Delta(\{u_n(\tau_j)\}_{1 \leq j \leq r})$ implies that extreme events situated far apart are almost independent. Of course, it is implied by strong mixing.

Suppose that $\Delta(\{u_n(\tau_j)\}_{1 \leq j \leq r})$ holds. A sequence of positive integers $(q_n)$ is said to be $\Delta(\{u_n(\tau_j)\}_{1 \leq j \leq r})$-separating if $q_n = o(n)$ and there exists a sequence $(l_n)$ such that $l_n = o(q_n)$ and $\lim_{n \to \infty} n q_n^{-1} \alpha_{n,l_n}(\tau_1,\ldots,\tau_r) = 0$.

We now present the technical conditions. The first one will be considered for "weak consistency" of the estimators.

CONDITION (C0). The stationary sequence $(X_n)$ has extremal index $\theta > 0$. $\Delta(u_n(\tau))$ holds for each $\tau > 0$ and there exists a probability measure $\pi = (\pi(i))_{i \geq 1}$, such that, for $i \geq 1$,

(C0.a) $\qquad \pi(i) = \lim_{n \to \infty} P(N_n^{(\tau)}(B_n) = i | N_n^{(\tau)}(B_n) > 0),$

with $B_n = (0, q_n/n]$, for some $\Delta(u_n(\tau))$-separating sequence $(q_n)$. Moreover, there exists a constant $\rho > 2$ such that, for each $\tau > 0$,

(C0.b) $$\sup_{n \geq 1} E(N_n^{(\tau)}(E))^\rho < \infty.$$

Condition (C0) ensures that the exceedance point process $N_n^{(\tau)}$ converges in distribution for every choice of $\tau > 0$ (see [24], Theorem 4.2). Let $0 < v < \rho$. Condition (C0.b) implies that $(N_n^{(\tau)}(E))^v$ are uniformly integrable and $\lim_{n \to \infty} E(N_n^{(\tau)}(E))^v = E(N_E^{(\tau)})^v < \infty$. In particular, the first and second moments of $N_E^{(\tau)}$ exist (see [4], page 338). They are given by $E(N_E^{(\tau)}) = \tau$ and $V(N_E^{(\tau)}) = \theta \tau E(\zeta_1)^2$.

The following set of conditions will be considered for characterizing the distributional asymptotics of the estimators.

CONDITION (C1). Condition (C0) holds. $\Delta(u_n(\tau_1), u_n(\tau_2))$ holds for each $\tau_1 > \tau_2 > 0$ and there exists a probability measure $\pi_2 = (\pi_2^{(\tau_2/\tau_1)}(i,j))_{i \geq j \geq 0, i \geq 1}$, such that, for $i \geq j \geq 0$, $i \geq 1$,

(C1.a) $\pi_2^{(\tau_2/\tau_1)}(i,j) = \lim_{n \to \infty} P(N_n^{(\tau_1)}(B_n) = i, N_n^{(\tau_2)}(B_n) = j | N_n^{(\tau_1)}(B_n) > 0),$

with $B_n = (0, q_n/n]$, for some $\Delta(u_n(\tau_1), u_n(\tau_2))$-separating sequence $(q_n)$.

Let us introduce the two-level exceedance point process $\mathbf{N}_n^{(\tau_1, \tau_2)} = (N_n^{(\tau_1)}, N_n^{(\tau_2)})$ for $\tau_1 > \tau_2 > 0$. Condition (C1) ensures that $\mathbf{N}_n^{(\tau_1, \tau_2)}$ converges in distribution to a point process with Laplace transform

$$E \exp\left(-\sum_{i=1}^{2} \int_E f_i \, dN^{(\tau_i)}\right) = \exp\left(-\tau_1 \theta \int_0^1 (1 - L(f_1(t), f_2(t))) \, dt\right),$$

where $N^{(\tau_i)}$ is the $i$th marginal of the limiting point process, $f_i \geq 0$ and $L$ is the Laplace transform of $\pi_2^{(\tau_2/\tau_1)}$ (see Theorem 2.5 in [33] and its proof). Let us denote by $(N_E^{(\tau_1)}, N_E^{(\tau_2)})$ the weak limit of $(N_n^{(\tau_1)}(E), N_n^{(\tau_2)}(E))$. By considering constant functions $f_i$, we deduce that

$$(N_E^{(\tau_1)}, N_E^{(\tau_2)}) \stackrel{d}{=} \left(\sum_{i=1}^{\eta(\theta \tau_1)} \zeta_{1,i}^{(\tau_2/\tau_1)}, \sum_{i=1}^{\eta(\theta \tau_1)} \zeta_{2,i}^{(\tau_2/\tau_1)}\right),$$

where $(\zeta_{1,i}^{(\tau_2/\tau_1)}, \zeta_{2,i}^{(\tau_2/\tau_1)})_{i \geq 1}$ is a sequence of i.i.d. integer vector r.v.s with distribution $\pi_2^{(\tau_2/\tau_1)}$ and $\eta(\theta \tau_1)$ is a r.v. with Poisson distribution and parameter $\theta\tau_1$ such that $\eta(\theta\tau_1)$ is independent of the $(\zeta_{1,i}^{(\tau_2/\tau_1)}, \zeta_{2,i}^{(\tau_2/\tau_1)})$ (see also Theorem 2 in [29]). The distribution $p_2^{(\tau_1,\tau_2)} = (p_2^{(\tau_1,\tau_2)}(i,j))_{i\geq j\geq 0}$ of $(N_E^{(\tau_1)}, N_E^{(\tau_2)})$ is given by

$$p_2^{(\tau_1,\tau_2)}(0,0) = P(\eta(\theta\tau_1) = 0) = e^{-\theta\tau_1},$$

$$p_2^{(\tau_1,\tau_2)}(i,j) = \sum_{k=1}^{i} P(\eta(\theta\tau_1) = k) P\left(\sum_{l=1}^{k} \zeta_{1,l}^{(\tau_2/\tau_1)} = i, \sum_{l=1}^{k} \zeta_{2,l}^{(\tau_2/\tau_1)} = j\right)$$

$$= e^{-\theta\tau_1} \sum_{k=1}^{i} \frac{(\theta\tau_1)^k}{k!} \pi_2^{(\tau_2/\tau_1),*k}(i,j), \qquad i \geq j \geq 0, i \geq 1,$$

where $\pi_2^{(\tau_2/\tau_1),*k}$ is the $k$th convolution of $\pi_2^{(\tau_2/\tau_1)}$, that is,

$$\pi_2^{(\tau_2/\tau_1),*k}(i,j)$$
$$= \begin{cases} 0, & i < k, \\ \displaystyle\sum_{\substack{i_1+\cdots+i_k=i \\ j_1+\cdots+j_k=j \\ i_q\geq j_q\geq 0, i_q\geq 1, 1\leq q\leq k}} \pi_2^{(\tau_2/\tau_1)}(i_1,j_1)\cdots\pi_2^{(\tau_2/\tau_1)}(i_k,j_k), & i \geq k. \end{cases}$$

Condition (C0.b) implies that $\mathrm{Cov}(N_E^{(\tau_1)}, N_E^{(\tau_2)}) = \theta\tau_1 E(\zeta_{1,1}^{(\tau_2/\tau_1)}\zeta_{2,1}^{(\tau_2/\tau_1)})$ is finite.

CONDITION (C2). Let $r > 2$ and $\phi > 0$. There exists a constant $D = D(r,\phi)$ such that, for $\phi \geq \tau_1 \geq \tau_2 \geq 0$,

(C2.a) $$\sup_{n\geq 1} E(N_n^{(\tau_1)}(E) - N_n^{(\tau_2)}(E))^r \leq D(\tau_1 - \tau_2).$$

Let $\theta_d \geq 3r/(r - (2+\mu))$, where $0 < \mu < ((r-2) \wedge 1/2)$. There exists a constant $C > 0$ such that, for every choice of $\tau_1 > \cdots > \tau_m > 0$, $m \geq 1$, $1 \leq l \leq n$,

(C2.b) $$\alpha_{n,l}(\tau_1,\ldots,\tau_m) \leq \alpha_l := Cl^{-\theta_d}.$$

$(r_n)$ is sequence such that $r_n \to \infty$ and $r_n = o(n)$ and there exists a sequence $(l_n)$ satisfying

(C2.c) $$l_n = o(r_n^{2/r}) \quad \text{and} \quad \lim_{n\to\infty} nr_n^{-1}\alpha_{l_n} = 0.$$

Note that condition (C2.a) provides an inequality which is quite natural to prove tightness criteria. Condition (C2.b) is satisfied by strong-mixing stationary sequences where the mixing coefficients vanish at least with a hyperbolic rate. The underlying idea to establish the asymptotic properties of the estimators is to split the block $I_j$ into a small block of length $l_n$ and a big block of length $r_n - l_n$. Condition (C2.c) ensures that $l_n$ is sufficiently large such that blocks that are not adjacent are asymptotically independent, but does not grow too fast such that the contributions of the small blocks are negligible.

Finally, we need a condition on the convergence rate of $r_n$ to infinity to guarantee that the extreme value approximations are sufficiently accurate.

CONDITION (C3). Let $m$ be an integer. The sequence $(r_n)$ satisfies

$$\lim_{n \to \infty} \sqrt{k_n}(\tau - r_n \bar{F}(u_{r_n}(\tau))) = 0$$

and

$$\lim_{n \to \infty} \sqrt{k_n} \sum_{l=1}^{m} |P(N_{r_n}^{(\tau)}(E) = l) - p^{(\tau)}(l)| = 0$$

locally uniformly for $\tau > 0$.

Note that, if $F$ is continuous, then $r_n \bar{F}(u_{r_n}(\tau)) = \tau$ and the first part of Condition (C3) is obviously satisfied. We now discuss the example of the first order stochastic equations with random coefficients. A special case is the squared ARCH(1) process introduced in [14]. This process is probably one of the most prominent financial time series model of the last two decades.

EXAMPLE 3.1. Let $X_0$ be a r.v. and let $(A_n, B_n)$, $n \geq 1$, be i.i.d. $(0, \infty)^2$-valued random vectors independent of $X_0$. Define $X_n$ by means of the stochastic difference equation

(3.1) $$X_n = A_n X_{n-1} + B_n, \qquad n \geq 1.$$

For sake of simplicity, we assume that the distribution of $(A_1, B_1)$ is absolutely continuous. Kesten [25] proved that there exists a r.v. $X$, independent of $(A_1, B_1)$, such that $X \stackrel{d}{=} A_1 X + B$. Assume that $X_0$ has the same distribution as $X$, so that $(X_n)$ is a strictly stationary sequence. According to Corollary 2.4.1 in [8], $(X_n)$ is also strongly mixing and absolutely regular with geometric rates.

Further, suppose that there exist $\kappa > 0$ and $\xi > 0$ such that

$$EA_1^\kappa = 1, \qquad E(A_1^\kappa \max(\log(A_1), 0)) < \infty,$$
$$EA_1^{\kappa+\xi} < \infty \quad \text{and} \quad EB_1^{\kappa+\xi} \in (0, \infty).$$

Under these moment assumptions, results of Goldie [17] show that there exit $c > 0$ and $\rho > 0$ such that

(3.2) $$\bar{F}(x) = cx^{-\kappa}(1 + O(x^{-\rho})), \qquad \text{as } x \to \infty.$$

We deduce that $u_n(\tau) = (cn/\tau)^{1/\kappa}(1 + O(n^{-\rho/\kappa}))$ as $n \to \infty$. The one-level point process of exceedances was studied in [19] and the multi-level point process of exceedances in [33].

Now we successively verify that our technical conditions hold. Let $R(x) = \sharp\{j \geq 1 \colon \tilde{X} \prod_{i=1}^{j} A_i > x\}$, where $P(\tilde{X} > x) = x^{-\kappa}$, $x \geq 1$, and define $\theta_k = P(R(1) = k)$, $k \geq 0$. Using results in [19] and in [33], we see that $\Delta(u_n(\tau))$ holds for each $\tau > 0$ and that $\theta = \theta_0$ and $\pi(k) = (\theta_{k-1} - \theta_k)/\theta_0$, $k \geq 1$, for any $(q_n)$ $\Delta(u_n(\tau))$-separating sequence such that $q_n = n^\varsigma$ with $0 < \varsigma < 1$. Moreover, by Lemma 6.1 with $\tau_1 = \tau$ and $\tau_2 = 0$, we deduce that $E(N_n^{(\tau)}(E))^3 < \infty$ and that Condition (C0) holds with $\rho = 3$.

By [33], $\Delta(u_n(\tau_1), u_n(\tau_2))$ holds for each $\tau_1 > \tau_2 > 0$ and

$$\theta \pi_2^{(\tau_2/\tau_1)}(i,j) = \left(P\left(R(1) = i-1, R\left(\left(\frac{\tau_1}{\tau_2}\right)^{1/\kappa}\right) = j\right)\right.$$
$$\left. - P\left(R(1) = i, R\left(\left(\frac{\tau_1}{\tau_2}\right)^{1/\kappa}\right) = j\right)\right)$$
$$+ \frac{\tau_2}{\tau_1}\left(P\left(R\left(\left(\frac{\tau_2}{\tau_1}\right)^{1/\kappa}\right) = i-1, R(1) = j-1\right)\right.$$
$$\left. - P\left(R\left(\left(\frac{\tau_2}{\tau_1}\right)^{1/\kappa}\right) = i-1, R(1) = j\right)\right)$$

for any $(q_n)$ $\Delta(u_n(\tau_1), u_n(\tau_2))$-separating sequence such that $q_n = n^\varsigma$ with $0 < \varsigma < 1$. Therefore, Condition (C1) holds.

By Lemma 6.1, $E(N_n^{(\tau_1)}(E) - N_n^{(\tau_2)}(E))^3 \leq K(\tau_1 - \tau_2)$ for $\phi \geq \tau_1 \geq \tau_2 \geq 0$. There exists a constant $C$ satisfying (C2.b) for any $\theta_d \geq 9/(1-\mu)$, where $0 < \mu < 1/2$ because $(X_n)$ is a geometrically strong-mixing sequence. Moreover, if $r_n = n^\varsigma$ with $0 < \varsigma < 1$ and $l_n = n^\gamma$ with $0 < \gamma < 2\varsigma/3$, then (C2.c) is satisfied. Therefore, Condition (C2) holds.

Under the assumptions on $(A_1, B_1)$, $F$ is absolutely continuous and $r_n \bar{F}(u_{r_n}(\tau)) = \tau$. Let us use Lemma 6.2 with $q_n = \lfloor n^\alpha \rfloor$, $m_n = \lfloor n^\beta \rfloor$, $\delta_n = \lfloor n^\gamma \rfloor$, $x_n = \lfloor n^\delta \rfloor$ and $r_n = \lfloor n^\varsigma \rfloor$ with $0 < \beta < \alpha < 1$, $0 < \gamma < \kappa^{-1}$, $\delta > 0$ and $0 < \varsigma < 1$, then there exists a constant $K$ such that

$$\sqrt{k_n} \sum_{l=1}^{m} |P(N_{r_n}^{(\tau)}(E) = l) - p^{(\tau)}(l)|$$
$$\leq K n^{(1-\zeta)/2}(n^{-\chi\zeta} + n^{(1-\alpha)\zeta} \eta^{2n^{\beta\zeta}/3} + \varphi^{n^{\alpha\zeta}} n^{\delta\zeta\xi})$$

locally uniformly for $\tau > 0$, with $\chi = (\alpha - \beta) \wedge (1-\alpha) \wedge \alpha \wedge \gamma \wedge \delta(\kappa - \epsilon) \wedge \rho/\kappa$, $0 < \eta < 1$, $0 < \varphi < 1$ and $0 < \epsilon < \kappa$. Finally, choose $1/(1+2\chi) < \zeta < 1$ such that Condition (C3) holds.

**4. Asymptotic properties of the estimators.** To characterize the asymptotic properties of the estimators, it is convenient to introduce $D^m_{\sigma,\phi} \equiv D([\sigma,\phi], R^m)$ [resp. $D^m \equiv D((0,\infty), R^m)$], the space of functions from $[\sigma,\phi]$ [resp. $(0,\infty)$] to $R^m$ which are càglàd (left-continuous with right-limits) equipped with the strong $J_1$-topology (see [44] where the spaces of càdlàg functions (right-continuous with left-limits) are equivalently considered). Let us recall that weak convergence (which will be denoted by $\Rightarrow$) in $D^m$ is equivalent to weak convergence of the restrictions of the stochastic processes to any compact $[\sigma,\phi]$, $0 < \sigma < \phi < \infty$.

We start this section by giving a "weak consistency" result.

PROPOSITION 4.1. *Suppose that (C0) holds. Let $(r_n)$ be a sequence such that $r_n \to \infty$ and $r_n = o(n)$, and $0 < \sigma < \phi < \infty$. Then*

$$(\hat{p}_n^{(\cdot)}(0), \ldots, \hat{p}_n^{(\cdot)}(m)) \Rightarrow (p^{(\cdot)}(0), \ldots, p^{(\cdot)}(m))$$

*in $D^{m+1}_{\sigma,\phi}$,*

$$(\hat{\pi}_n^{(\cdot)}(1), \ldots, \hat{\pi}_n^{(\cdot)}(m)) \Rightarrow (\pi(1), \ldots, \pi(m))$$

*in $D^m_{\sigma,\phi}$,*

$$(\hat{\theta}_{1,n}^{(\cdot)}, \hat{\theta}_{2,n}^{(\cdot)}(m), \hat{\theta}_{3,n}^{(\cdot)}(m)) \Rightarrow (\theta, \theta_2(m), \theta_3^{(\cdot)}(m))$$

*in $D^3_{\sigma,\phi}$,*

$$\hat{\bar{\pi}}_n(m) \xrightarrow{P} \pi(m), \qquad m \geq 1 \quad and \quad \hat{\bar{\theta}}_{1,n} \xrightarrow{P} \theta.$$

We continue with a series of results leading to a characterization of the distributional asymptotics of the estimators of the limiting cluster size probabilities. We first introduce the following centered processes:

$$e_{j,n}(\cdot) = \sqrt{k_n}(p_n^{(\cdot)}(j) - P(N_{r_n,1}^{(\cdot)} = j)), \qquad j \geq 0,$$
$$\bar{e}_n(\cdot) = \sqrt{k_n}(\bar{p}_n^{(\cdot)} - r_n P(X_1 > u_{r_n}(\cdot))),$$

where

$$\bar{p}_n^{(\tau)} = \sum_{i=1}^{\infty} i p_n^{(\tau)}(i) = \frac{1}{k_n} \sum_{j=1}^{k_n} N_{r_n,j}^{(\tau)} = \frac{1}{k_n} \sum_{i=1}^{r_n k_n} 1_{\{X_i > u_{r_n}(\tau)\}}.$$

$\bar{p}_n^{(\cdot)}$ is called the tail empirical distribution and $\bar{e}_n(\cdot)$ the tail empirical process. They are very useful tools for studying the asymptotic properties of tail index estimators (see, e.g., [9, 34]) or for inference of multivariate extreme value distributions [18]. The weak convergence of the tail empirical process of strong-mixing (resp. absolute regular) stationary sequences has been studied by [37] (resp. by [37], [10] and [11]). Note that the absolute regularity condition implies the strong-mixing condition which implies $\Delta(\{u_n(\tau_j)\}_{1 \leq j \leq r})$ for every choice of $\tau_1 > \cdots > \tau_r > 0$, $r \geq 1$. The following theorem deals with the weak convergence of the process

$$E_{m,n}(\cdot) = (e_{0,n}(\cdot), \ldots, e_{m,n}(\cdot), \bar{e}_n(\cdot))$$

in $D^{m+2}$. It will be useful throughout this section.

THEOREM 4.1. *Suppose that (C1) and (C2) hold. There exists a pathwise continuous centered Gaussian process*

$$E_m(\cdot) = (e_0(\cdot), \ldots, e_m(\cdot), \bar{e}(\cdot))$$

*with covariance functions defined for $0 < \tau_2 \leq \tau_1$ by:*

- *if $i = 0, \ldots, m$,*

$$\text{cov}(e_i(\tau_1), e_i(\tau_2)) = p_2^{(\tau_1, \tau_2)}(i, i) - p^{(\tau_1)}(i)p^{(\tau_2)}(i),$$

$$\text{cov}(e_i(\tau_1), \bar{e}(\tau_2)) = \sum_{j=0}^{i} j p_2^{(\tau_1, \tau_2)}(i, j) - \tau_2 p^{(\tau_1)}(i),$$

$$\text{cov}(\bar{e}(\tau_1), e_i(\tau_2)) = \sum_{j=i}^{\infty} j p_2^{(\tau_1, \tau_2)}(j, i) - \tau_1 p^{(\tau_2)}(i),$$

- *if $0 \leq i < j \leq m$,*

$$\text{cov}(e_i(\tau_1), e_j(\tau_2)) = -p^{(\tau_1)}(i)p^{(\tau_2)}(j),$$

- *if $0 \leq j < i \leq m$,*

$$\text{cov}(e_i(\tau_1), e_j(\tau_2)) = p_2^{(\tau_1, \tau_2)}(i, j) - p^{(\tau_1)}(i)p^{(\tau_2)}(i),$$

- *and*

$$\text{cov}(\bar{e}(\tau_1), \bar{e}(\tau_2)) = -\ln(p^{(\tau_1)}(0)) \sum_{0 \leq j \leq i, 1 \leq i} ij \pi_2^{(\tau_2/\tau_1)}(i, j),$$

*such that $E_{m,n} \Rightarrow E_m$ in $D^{m+2}$.*

Let us compare the conditions in [37] that are needed for convergence of $\bar{e}(\cdot)$ in the case of strong-mixing sequences with our conditions. First we have to impose that the threshold, $u_n$, in [37], is such that $u_n = O(u_{r_n}(\tau))$. Then Condition C1 in [37] is equivalent to our condition (C2.a). Condition D2 in [37] is slightly weaker than our condition (C2.b) and condition (C2.c) since we also assume that $l_n = o(r_n^{2/r})$. Condition C3 in [37] is implied by our Condition (C1), but it appears as a natural sufficient condition when $u_n = O(u_{r_n}(\tau))$.

Now let us consider the estimators of the compound probabilities and introduce the following processes:

$$\hat{e}_{j,n}(\cdot) = \sqrt{k_n}(\hat{p}_n^{(\cdot)}(j) - p^{(\cdot)}(j)), \qquad j \geq 0.$$

THEOREM 4.2. *Suppose that (C1), (C2) and (C3) hold. Let $0 < \sigma < \phi < \infty$. Then*

$$(\hat{e}_{0,n}(\cdot), \ldots, \hat{e}_{m,n}(\cdot)) \Rightarrow (\hat{e}_0(\cdot), \ldots, \hat{e}_m(\cdot))$$

*in $D_{\sigma,\phi}^{m+1}$, where*

$$\hat{e}_j(\cdot) = e_j(\cdot) - h_j(\cdot)\bar{e}(\cdot)$$

*and $h_j(\tau) = \partial p^{(\bar{\tau})}(j)/\partial \bar{\tau}|_{\bar{\tau}=\tau}$.*

Note that the $h_j(\cdot)$ satisfy the recursion

$$h_0(\cdot) = p^{(\cdot)}(0) \ln p^{(1)}(0),$$

$$h_j(\cdot) = -\frac{\ln(p^{(\cdot)}(0))}{j} \sum_{i=1}^{j} i\pi(i) \left( \frac{\ln(p^{(1)}(0))}{\ln(p^{(\cdot)}(0))} p^{(\cdot)}(j-i) + h_{j-i}(\cdot) \right), \qquad j \geq 1.$$

In order to address the asymptotic properties of the estimators of the limiting cluster size probabilities, we construct several processes. Following [6], we define recursively the processes $\hat{d}_j(\cdot)$ using the intermediate processes

$$w_j(\cdot) := \frac{p^{(\cdot)}(j)}{(\ln(p^{(\cdot)}(0))p^{(\cdot)}(0))^2} \hat{e}_0(\cdot) - \frac{1}{jp^{(\cdot)}(0)} \sum_{i=0}^{j-1}(j-i)\pi(j-i)\hat{e}_i(\cdot)$$

$$- \frac{1}{\ln(p^{(\cdot)}(0))p^{(\cdot)}(0)} \hat{e}_j(\cdot) - \frac{1}{jp^{(\cdot)}(0)} \sum_{i=1}^{j-1} ip^{(\cdot)}(j-i)\hat{d}_i(\cdot)$$

by $\hat{d}_0(\cdot) = -\hat{e}_0(\cdot)/p^{(\cdot)}(0)$ and for $j \geq 1$,

$$\hat{d}_j(\cdot) := \begin{cases} w_j(\cdot), & \text{if } \pi(j) > 0 \text{ and } \sum_{i=1}^{j}\pi(i) < 1, \\ \min\left\{w_j(\cdot), -\sum_{i=1}^{j-1}\hat{d}_i(\cdot)\right\}, & \text{if } \pi(j) > 0 \text{ and } \sum_{i=1}^{j}\pi(i) = 1, \\ \max\{0, w_j(\cdot)\}, & \text{if } \pi(j) = 0 \text{ and } \sum_{i=1}^{j}\pi(i) < 1, \\ \max\left\{0, \min\left\{w_j(\cdot), -\sum_{i=1}^{j-1}\hat{d}_i(\cdot)\right\}\right\}, & \text{if } \pi(j) = 0 \text{ and } \sum_{i=1}^{j}\pi(i) = 1. \end{cases}$$

Note that the process $\hat{d}_j(\cdot)$ depends on the support of the limiting cluster size distribution. It is not in general a Gaussian process because of the truncations in its construction, except if $\pi(i) > 0$ for $i = 1, \ldots, j$ and $\sum_{i=1}^{j} \pi(i) < 1$.

In the following corollary we derive the weak convergence of the processes

$$\hat{d}_{j,n}(\cdot) = \sqrt{k_n}(\hat{\pi}_n^{(\cdot)}(j) - \pi(j)), \qquad j \geq 1,$$

and the asymptotic behavior of

$$\bar{d}_{j,n} = \sqrt{k_n}(\widehat{\bar{\pi}}_n(j) - \pi(j)), \qquad j \geq 1.$$

COROLLARY 4.1. *Suppose that (C1), (C2) and (C3) hold. Let $0 < \sigma < \phi < \infty$. Then*

$$(\hat{d}_{1,n}(\cdot), \ldots, \hat{d}_{m,n}(\cdot)) \Rightarrow (\hat{d}_1(\cdot), \ldots, \hat{d}_m(\cdot))$$

*in $D_{\sigma,\phi}^m$ and*

$$(\bar{d}_{1,n}, \ldots, \bar{d}_{m,n}) \xrightarrow{d} \left( \frac{1}{\phi - \sigma} \int_\sigma^\phi \hat{d}_1(\tau) \, d\tau, \ldots, \frac{1}{\phi - \sigma} \int_\sigma^\phi \hat{d}_m(\tau) \, d\tau \right).$$

We end this section by focusing on the estimators of the extremal index.

COROLLARY 4.2. *Suppose that (C1), (C2) and (C3) hold. Let $0 < \sigma < \phi < \infty$. Then*

$$\sqrt{k_n}(\hat{\theta}_{1,n}^{(\cdot)} - \theta) \Rightarrow -\frac{1}{(\cdot)p^{(\cdot)}(0)}\hat{e}_0(\cdot)$$

*in $D_{\sigma,\phi}^1$,*

$$\sqrt{k_n}(\hat{\theta}_{2,n}^{(\cdot)}(m) - \theta_2(m)) \Rightarrow -(\theta_2(m))^2 \sum_{j=1}^{m} j\hat{d}_j(\cdot)$$

*in $D_{\sigma,\phi}^1$,*

$$\sqrt{k_n}(\hat{\theta}_{3,n}^{(\cdot)}(m) - \theta_3^{(\cdot)}(m)) \Rightarrow \frac{(\sum_{j=0}^{m}(j-(\cdot))^2 \hat{e}_j(\cdot) - \theta_3^{(\cdot)}(m) \sum_{j=1}^{m} j^2 \hat{d}_j(\cdot))}{(\cdot) \sum_{j=1}^{m} j^2 \pi(j)}$$

*in $D_{\sigma,\phi}^1$ and*

$$\sqrt{k_n}(\widehat{\bar{\theta}}_{1,n} - \theta) \xrightarrow{d} -\frac{1}{\phi - \sigma} \int_\sigma^\phi \frac{1}{\tau p^{(\tau)}(0)} \hat{e}_0(\tau) \, d\tau.$$

Note that the asymptotic variance of $\hat{\theta}_{1,n}^{(\tau)}$ is given by

$$\tau^{-2}\left(e^{\theta\tau} - 2\theta\tau - 1 + \theta^3\tau \sum_{j=1}^{\infty} j^2 \pi(j)\right).$$

It can be estimated by using the estimators of the limiting cluster probabilities $\hat{\pi}_n^{(\tau)}(j)$ and the estimator of the extremal index $\hat{\theta}_{1,n}^{(\tau)}$.

**5. Simulation study.** A simulation study is conducted to investigate the performance of the estimators on large samples and to make a comparison with existing estimators.

(i) *Performance on large samples.* Data are simulated from three stationary Markov processes:

- a squared ARCH(1) process: $X_n = (\eta + \lambda X_{n-1})Z_n^2$, $n \geq 2$, where $Z_n$ are i.i.d. standard Gaussian r.v.s, $\eta = 2 \times 10^{-5}$, $\lambda = 0.5$ and $X_1$ is a r.v. drawn from the stationary distribution of the chain. The limiting cluster size probabilities and the extremal index have been computed by simulations in [19]: $\pi(1) = 0.751$, $\pi(2) = 0.168$, $\pi(3) = 0.055$, $\pi(4) = 0.014$, $\pi(5) = 0.008$, $\theta = 0.727$.
- a max-AR(1) process: $X_n = \max\{(1-\theta)X_{n-1}, W_n\}$, $n \geq 2$, where $W_n$ are i.i.d. unit Fréchet r.v.s, $\theta = 0.5$ and $X_1 = W_1/\theta$. By [33], $\pi(1) = 0.5$, $\pi(2) = 0.25$, $\pi(3) = 0.125$, $\pi(4) = 0.0625$, $\pi(5) = 0.031$, $\theta = 0.5$.
- an AR(1) process with uniform marginal: $X_n = r^{-1} X_{n-1} + \varepsilon_n$, $n \geq 2$, where $(\varepsilon_n)$ are i.i.d. r.v.s uniformly distributed on $\{0, 1/r, \ldots, (r-1)/r\}$, $r = 4$ and $X_1$ is uniformly distributed on $(0, 1)$. By [33], $\pi(1) = 0.75$, $\pi(2) = 0.1875$, $\pi(3) = 0.0469$, $\pi(4) = 0.0117$, $\pi(5) = 0.0029$, $\theta = 0.75$.

To compare the performance of the estimators, 500 sequences of length $n = 2000$ were simulated from the three processes. We have considered the ratios $\hat{\pi}_n^{(1)}(j)/\pi(j)$ for $j = 1, \ldots, 5$, $\hat{\theta}_{1,n}^{(1)}/\theta$, and $\hat{\theta}_{j,n}^{(1)}(m)/\theta$ for $j = 2, 3$ and $m = 8$. The graphs show the average over the 500 samples.

In Figures 1 and 2 the means and the root mean squared errors (RMSE) of the ratios are plotted as a function of $k_n$. The bias of $\hat{\pi}_n^{(1)}(1)$ is small and approximatively stable with respect to $k_n$ for the three processes. The biases of $\hat{\pi}_n^{(1)}(2)$ and $\hat{\pi}_n^{(1)}(3)$ are small for the squared ARCH(1) process and the max-AR(1) process but large for the AR(1) process.

For $j \geq 4$, the biases of the estimators can be relatively large and it seems very difficult to have good estimates of $\pi(j)$ in the case of a data set of length 2000. The RMSE of the ratios increase dramatically with $j$ because of the biases. Note also that a minimum of the RMSE with respect to $k_n$ can not always be found. An optimal choice of $k_n$ based on the RMSE criterion will depend on the process and on the limiting cluster size probabilities.

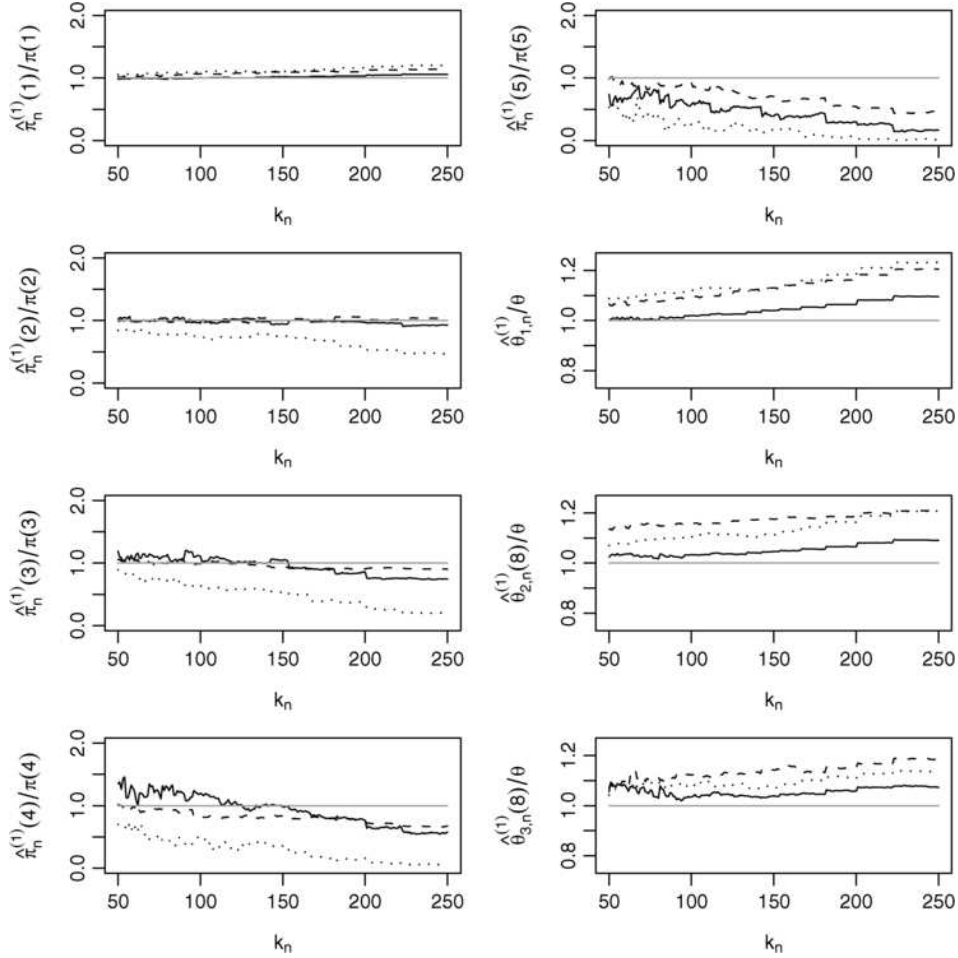

FIG. 1. *Means of the ratios of the cluster size probabilities* $\hat{\pi}_n^{(1)}(1)/\pi(1)$, $\hat{\pi}_n^{(1)}(2)/\pi(2)$, $\hat{\pi}_n^{(1)}(3)/\pi(3)$, $\hat{\pi}_n^{(1)}(4)/\pi(4)$, $\hat{\pi}_n^{(1)}(5)/\pi(5)$, *and means of the ratios of the extremal index* $\hat{\theta}_{1,n}^{(1)}/\theta$, $\hat{\theta}_{2,n}^{(1)}(8)/\theta$ *and* $\hat{\theta}_{3,n}^{(1)}(8)/\theta$ *as a function of* $k_n = 50, \ldots, 250$ *for the squared* ARCH(1) *process (——), the max-*AR(1) *process (- - - -) and the* AR(1) *process (· · · ·). The graphs show the average over 500 samples of length* $n = 2000$.

The bias of $\hat{\theta}_{1,n}^{(1)}$ is lower than those of $\hat{\theta}_{2,n}^{(1)}(m)$ and $\hat{\theta}_{3,n}^{(1)}(m)$ for the squared ARCH(1) process and the max-AR(1) process. But for the AR(1) process, the bias of $\hat{\theta}_{3,n}^{(1)}(m)$ is the smallest. $\hat{\theta}_{1,n}^{(1)}$ and $\hat{\theta}_{2,n}^{(1)}(m)$ perform in the same way in terms of RMSE and better than $\hat{\theta}_{3,n}^{(1)}(m)$.

(ii) *Comparison with existing estimators on large samples.* Data are simulated from the squared ARCH(1) process defined below. 500 sequences of length $n = 2000$ were also used. For the limiting cluster probabilities com-parisons are made between $\dot{\pi}_n(i) = \hat{\pi}_n^{(1)}(j)$, $\widehat{\tilde{\pi}}_n(j)$ with $\sigma = 0.7$ and $\phi = 1.3$, Hsing's estimators $\hat{\pi}_{n,1}(j)$ with $n/s_n = k_n/2$ and Ferro's estimators $\tilde{\pi}_n(j)$ with $N = k_n$ (see [15], equation (4.12)). For the extremal index comparisons are made between $\dot{\theta}_n = \hat{\theta}_{1,n}^{(1)}$, $\widehat{\tilde{\theta}}_{1,n}$ with $\sigma = 0.7$ and $\phi = 1.3$, Ferro and Segers' estimator $\tilde{\theta}_n(u)$ with $u = X_{n-k_n+1:n}$ (see [16], equation (5)), Hsing's estimator $\tilde{\theta}_n$ with $n/s_n = k_n/2$ (see [21], page 137) and the runs estima-tor $\hat{\theta}_n^R(p, u)$ with $p = \lfloor r_n/6 \rfloor$, $u = X_{n-\lfloor n/s_n \rfloor:n}$ and $n/s_n = k_n/2$ (see [43], page 282).

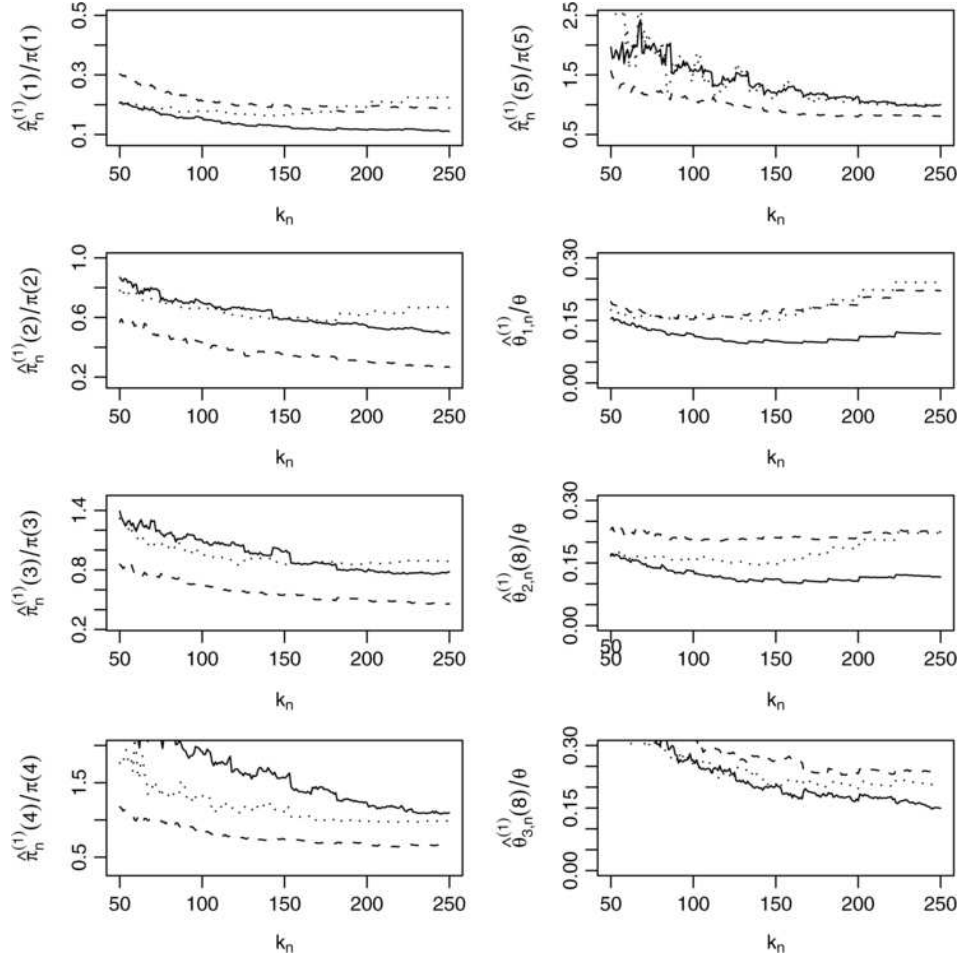

Fig. 2. *RMSE of the ratios of the cluster size probabilities $\hat{\pi}_n^{(1)}(1)/\pi(1)$, $\hat{\pi}_n^{(1)}(2)/\pi(2)$, $\hat{\pi}_n^{(1)}(3)/\pi(3)$, $\hat{\pi}_n^{(1)}(4)/\pi(4)$, $\hat{\pi}_n^{(1)}(5)/\pi(5)$, and RMSE of the ratios of the extremal index $\hat{\theta}_{1,n}^{(1)}/\theta$, $\hat{\theta}_{2,n}^{(1)}(8)/\theta$ and $\hat{\theta}_{3,n}^{(1)}(8)/\theta$ as a function of $k_n = 50, \ldots, 250$ for the squared $\mathrm{ARCH}(1)$ process (———), the max-$\mathrm{AR}(1)$ process (- - - -) and the $\mathrm{AR}(1)$ process ($\cdots$). The graphs show the average over 500 samples of length $n = 2000$.*

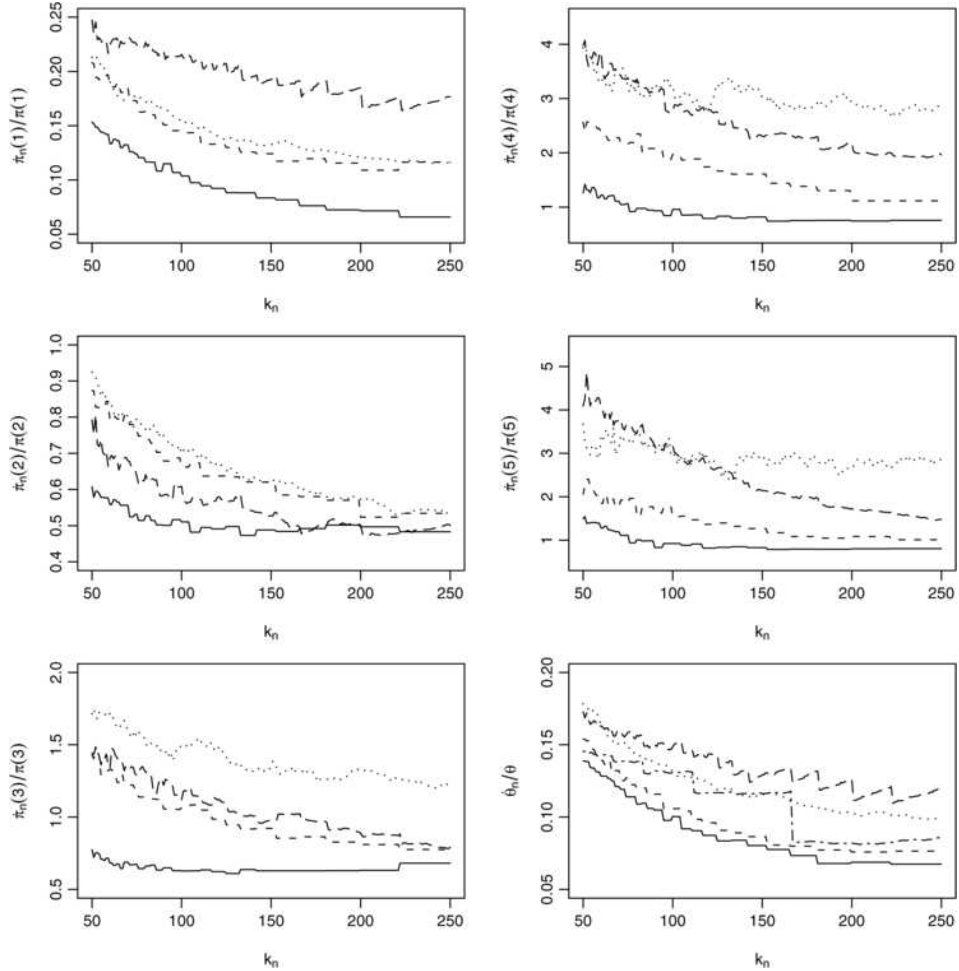

FIG. 3. *RMSE of the ratios of the cluster size probabilities $\dot{\pi}_n(1)/\pi(1)$, $\dot{\pi}_n(2)/\pi(2)$, $\dot{\pi}_n(3)/\pi(3)$, $\dot{\pi}_n(4)/\pi(4)$, $\dot{\pi}_n(5)/\pi(5)$ as a function of $k_n = 50, \ldots, 250$, for $\dot{\pi}_n = \hat{\pi}_n^{(1)}$ (- - - -), $\dot{\pi}_n = \widehat{\bar{\pi}}_n$ (———), $\dot{\pi}_n = \tilde{\pi}_n$ (Ferro's estimators · · · ·) and $\dot{\pi}_n = \hat{\pi}_{n,1}$ (Hsing's estimator – – –). RMSE of the ratios of the extremal index $\dot{\theta}_n/\theta$ as a function of $k_n = 50, \ldots, 250$, for $\dot{\theta}_n = \hat{\theta}_{1,n}^{(1)}$ (- - - -), $\dot{\theta}_n = \widehat{\bar{\theta}}_{1,n}$ (———), $\dot{\theta}_n = \tilde{\theta}_n$ (Ferro and Segers' estimator · · · ·), $\dot{\theta}_n = \bar{\theta}_n$ (Hsing's estimator – – –) and $\dot{\theta}_n = \hat{\theta}_n^R$ (runs estimator - – - -).*

In Figure 3 the RMSE of the ratios $\dot{\pi}_n(i)/\pi(i)$ for $i = 1, \ldots, 5$, and $\dot{\theta}_n/\theta$ are plotted. For the limiting cluster size probabilities and the extremal index, the smoothed versions $\widehat{\bar{\pi}}_n$ and $\widehat{\bar{\theta}}_{1,n}$ perform uniformly better than the unsmoothed estimators $\hat{\pi}_n^{(1)}$ and $\hat{\theta}_{1,n}^{(1)}$ which perform uniformly better than the other estimators [except for $\pi(2)$, where Hsing's estimator should be preferred to the unsmoothed estimator]. As Ferro and Segers' estimators, our estimators only require the choice of a sequence, but their performance is more favorable.

**6. Proofs.** Throughout we let $K$ be a generic constant whose value may change from line to line.

LEMMA 6.1. *Consider the first order stochastic equation with random coefficients of Example 3.1. There exists a constant $K$ such that, for $\phi \geq \tau_1 \geq \tau_2 \geq 0$ and $n \geq 1$,*

$$E(N_n^{(\tau_1)}(E) - N_n^{(\tau_2)}(E))^3 \leq K(\tau_1 - \tau_2).$$

PROOF. Let $I_n(\tau_1, \tau_2) = (u_n(\tau_1), u_n(\tau_2)]$. By using the same arguments as in the proof of Lemma 4.1 in [10], we can show that there exists a constant $K$ such that, for $\phi \geq \tau_1 \geq \tau_2 \geq 0$ and $n \geq 1$,

$$c_{i,j} = P(X_j \in I_n(\tau_1, \tau_2) | X_i \in I_n(\tau_1, \tau_2)) \leq K\left(\frac{1}{n} + \varphi^{j-i}\right), \qquad j \geq i \geq 1,$$

where $\varphi = EA_1^\xi < 1$ for $\xi \in (0, \kappa)$. By the stationary and Markov property, we get

$$E(N_n^{(\tau_1)}(E) - N_n^{(\tau_2)}(E))^3$$
$$\leq 3! n \sum_{i,j \geq 1, i+j \leq n+1} E 1_{\{X_1 \in I_n(\tau_1,\tau_2)\}} 1_{\{X_i \in I_n(\tau_1,\tau_2)\}} 1_{\{X_{i+j-1} \in I_n(\tau_1,\tau_2)\}}$$
$$\leq 3!(\tau_1 - \tau_2) \sum_{i,j \geq 1, i+j \leq n+1} c_{1,i} c_{i,i+j-1}$$
$$\leq 3!(\tau_1 - \tau_2) K^2 \sum_{i,j \geq 1, i+j \leq n+1} \left(\frac{1}{n} + \varphi^{j-1}\right)\left(\frac{1}{n} + \varphi^{i-1}\right)$$
$$\leq K(\tau_1 - \tau_2). \qquad \square$$

LEMMA 6.2. *Consider the first order stochastic equation with random coefficients of Example 3.1. Let $(q_n)$, $(m_n)$, $(\delta_n)$, $(x_n)$ be sequences of integers such that $q_n \to \infty$ and $q_n = o(n)$, $m_n \to \infty$ and $m_n = o(q_n)$, $\delta_n \to \infty$ and $n\delta_n^{-\kappa} \to \infty$ and $x_n \to \infty$ as $n \to \infty$. Then for each $l \geq 0$, there exists a constant $K$ such that*

$$|P(N_n^{(\tau)}(E) = l) - p^{(\tau)}(l)| \leq K\left(\frac{m_n}{q_n} + \frac{q_n}{n} + \frac{1}{q_n} + \frac{n}{q_n}\eta^{2m_n/3} + \frac{q_{\lfloor n\delta_n^{-\kappa} \rfloor}}{n\delta_n^{-\kappa}} \right.$$
$$\left. + \delta_n^{-1} + \varphi^{q_n} x_n^\xi + x_n^{-(\kappa-\epsilon)} + n^{-\rho/\kappa}\right)$$

CLUSTER SIZE DISTRIBUTION OF EXTREME VALUES 21

*locally uniformly for $\tau > 0$, where $0 < \eta < 1$, $0 < \varphi < 1$, $0 < \xi < \kappa$ and $0 < \epsilon < \kappa$.*

PROOF. Write $d(n,l) = |P(N_n^{(\tau)}(E) = l) - p^{(\tau)}(l)|$. Let $\theta_n^{(\tau)} = n \times P(N_n^{(\tau)}(B_n) > 0)/(\tau q_n)$, where $B_n = (0; q_n/n]$. Let $\zeta_{i,n}^{(\tau)}$, $i \geq 1$, be i.i.d. integer-valued r.v.s such that

$$P(\zeta_{i,n}^{(\tau)} = m) = P(N_n^{(\tau)}(B_n) = m | N_n^{(\tau)}(B_n) > 0), \qquad m \geq 1,$$

and $\eta(\theta_n^{(\tau)}\tau)$ be a Poisson r.v. with parameter $\theta_n^{(\tau)}\tau$ and independent of the $\zeta_{i,n}^{(\tau)}$. We have that

$$d(n,l) \leq \left| P(N_n^{(\tau)}(E) = l) - P\left(\sum_{i=1}^{\eta(\theta_n^{(\tau)}\tau)} \zeta_{i,n}^{(\tau)} = l\right) \right|$$

$$+ \left| P\left(\sum_{i=1}^{\eta(\theta_n^{(\tau)}\tau)} \zeta_{i,n}^{(\tau)} = l\right) - p^{(\tau)}(l) \right|$$

$$=: \mathrm{I}_l + \mathrm{II}_l.$$

By using Theorem 2 in [30], we deduce that

$$\mathrm{I}_l \leq 2\tau \frac{m_n}{q_n} + 2\tau \frac{q_n}{n} + \frac{n}{q_n} \min\{6\alpha_{n,m_n}^{2/3}(\tau); \beta_{n,m_n}(\tau)\},$$

where

$$\beta_{n,l}(\tau) \equiv \sup_{1 \leq t \leq n-l} E \sup |P(B|\mathcal{F}_{1,t}^{(\tau)}) - P(B) : B \in \mathcal{F}_{t+l,n}^{(\tau)}|.$$

Note that, since $(X_n)$ is a geometrically absolute regular sequence, there exists a constant $0 < \eta < 1$ such that, for every choice of $\tau > 0$ and $1 \leq l \leq n$, $\alpha_{n,l}(\tau) \leq \beta_{n,l}(\tau) \leq O(\eta^l)$.

By (1.3) and (1.4), we deduce that there exist constants $K_{1,l} > 0$ and $K_{2,l} > 0$ such that, locally uniformly for $\tau > 0$,

$$\mathrm{II}_l \leq K_{1,l}|\theta_n^{(\tau)} - \theta_0| + K_{2,l} \sum_{k=1}^{l} |\pi_n^{(\tau)}(k) - \pi(k)|$$

$$\leq K_{1,l}|\theta_n^{(\tau)} - \theta_0| + 2\theta_0^{-1} K_{2,l} \sum_{k=1}^{l+1} \left| \theta_0 \sum_{j=k}^{q_n} \pi_n^{(\tau)}(j) - \theta_{k-1} \right|.$$

Let $\theta_{k,n}^{(\tau)} = P(N_n^{(\tau)}(B_n) = k | X_0 > u_n(\tau))$. Note that

$$|\theta_n^{(\tau)} - \theta_0| \leq |\theta_n^{(\tau)} - \theta_{0,n}^{(\tau)}| + |\theta_{0,n}^{(\tau)} - \theta_0| =: \mathrm{II}a + \mathrm{II}b$$

and for $k \geq 1$,

$$\left|\theta_0 \sum_{j=k}^{q_n} \pi_n^{(\tau)}(j) - \theta_{k-1}\right| \leq \left|\theta_0 \frac{P(N_n^{(\tau)}(B_n) \geq k)}{P(N_n^{(\tau)}(B_n) \geq 1)} - \frac{\theta_0}{\theta_n^{(\tau)}} \theta_{k-1,n}^{(\tau)}\right| + \left|\frac{\theta_0}{\theta_n^{(\tau)}} - 1\right| \theta_{k-1,n}^{(\tau)}$$

$$+ |\theta_{k-1,n}^{(\tau)} - \theta_{k-1}| =: \mathrm{II}c_k + \mathrm{II}d_k + \mathrm{II}e_k.$$

By using the same arguments as for the proof of Lemma 2.4 in [33], we have for $k \geq 1$

$$|P(N_n^{(\tau)}(B_n) \geq k) - (q_n - k + 1)P(N_n^{(\tau)}(B_n) = k - 1, X_0 > u_n(\tau))|$$
$$\leq k P(M_{0,q_n} > u_n(\tau), M_{q_n,2q_n} > u_n(\tau)),$$

where $M_{i,j} = \max\{X_l : l = i+1, \ldots, j\}$. It follows that for $k \geq 1$

$$\mathrm{II}c_k \leq k \frac{\theta_0}{\theta_n^{(\tau)}} \frac{nP(M_{0,q_n} > u_n(\tau), M_{q_n,2q_n} > u_n(\tau))}{\tau q_n} + \frac{k-1}{q_n} \frac{\theta_0}{\theta_n^{(\tau)}}$$

and

$$\mathrm{II}a \leq nP(M_{0,q_n} > u_n(\tau), M_{q_n,2q_n} > u_n(\tau))/(\tau q_n).$$

Now observe that

$$P(M_{0,q_n} > u_n(\tau), M_{q_n,2q_n} > u_n(\tau))$$
$$= P(\{\{M_{0,q_n-m_n} > u_n(\tau)\} \cup \{M_{q_n-m_n,q_n} > u_n(\tau)\}\} \cap \{M_{q_n,2q_n} > u_n(\tau)\})$$
$$\leq P(M_{q_n-m_n,q_n} > u_n(\tau)) + \alpha_{n,m_n}(\tau) + P^2(M_{0,q_n} > u_n(\tau))$$
$$\leq \tau \frac{m_n}{n} + \alpha_{n,m_n}(\tau) + \left(\tau \frac{q_n}{n}\right)^2 \theta_0^2$$

and, therefore,

$$\mathrm{II}a + \sum_{k=1}^{l+1}(\mathrm{II}c_k + \mathrm{II}d_k) \leq K\left(\frac{m_n}{q_n} + \frac{n}{q_n}\alpha_{n,m_n}(\tau) + \frac{q_n}{n} + \frac{1}{q_n}\right).$$

Let $\sigma_k = \sum_{j=k}^{\infty} \theta_j = P(R(1) \geq k) = \int_1^{\infty} P(\sharp\{j \geq 1 : \prod_{i=1}^{j} A_i > x^{-1}\} \geq k)\kappa \times x^{-\kappa-1} dx$. We have that $\theta_{k-1} = \sigma_k - \sigma_{k-1}$ for $k \geq 1$. Then

$$\mathrm{II}b \leq \left|\sum_{j=0}^{q_n} \theta_{j,n}^{(\tau)} - \sigma_0\right| + \left|\sum_{j=1}^{q_n} \theta_{j,n}^{(\tau)} - \sigma_1\right|$$

$$\mathrm{II}e_k \leq \left|\sum_{j=k-1}^{q_n} \theta_{j,n}^{(\tau)} - \sigma_{k-1}\right| + \left|\sum_{j=k}^{q_n} \theta_{j,n}^{(\tau)} - \sigma_k\right|, \quad k \geq 1.$$

Let us define the probability measure, $Q_n$, on $(1, \infty)$ by

$$Q_n(dx) = P((u_n(\tau))^{-1} X_0 \in dx)/P((u_n(\tau))^{-1} X_0 > 1).$$

As in [19], we introduce the process $(\Delta_n)$ defined by $\Delta_0 = 0$ and $\Delta_n = A_n \Delta_{n-1} + B_n$, $n \geq 1$. We have that $\Delta_n \geq 0$ and $X_n = X_0 \prod_{i=1}^n A_i + \Delta_n$ for $n \geq 1$. Let

$$B_k(q_n, (\Delta_j)_{j=1,\ldots,q_n}, Q_n)$$
$$= \int_1^\infty P\left(\sharp\left\{1 \leq j \leq q_n : X_0 \prod_{i=1}^j A_i + \Delta_j > u_n(\tau)\right\}\right.$$
$$\left.\geq k \Big| (u_n(\tau))^{-1} X_0 = x\right) Q_n(dx).$$

Note that $P(N_n^{(\tau)}(B_n) \geq k | X_0 > u_n(\tau)) = B_k(q_n, (\Delta_j)_{j=1,\ldots,q_n}, Q_n)$ and that

$$\left|\sum_{j=k}^{q_n} \theta_{j,n}^{(\tau)} - \sigma_k\right| \leq |B_k(q_n, (\Delta_j)_{j=1,\ldots,q_n}, Q_n) - B_k(q_n, (0)_{j=1,\ldots,q_n}, Q_n)|$$
$$+ |B_k(q_n, (0)_{j=1,\ldots,q_n}, Q_n) - B_k(\infty, (0)_{j=1,\ldots,\infty}, Q_n)|$$
$$+ |B_k(\infty, (0)_{j=1,\ldots,\infty}, Q_n) - B_k(\infty, (0)_{j=1,\ldots,\infty}, Q)|.$$

We now consider successively each term of the upper bound:

(i) On the one hand, we have that

$$\int_1^\infty P\left(\sharp\left\{1 \leq j \leq q_n : X_0 \prod_{i=1}^j A_i + \Delta_j > u_n(\tau)\right\} \geq k \Big| (u_n(\tau))^{-1} X_0 = x\right)$$
$$\times Q_n(dx)$$
$$\geq \int_1^\infty P\left(\sharp\left\{1 \leq j \leq q_n : X_0 \prod_{i=1}^j A_i > u_n(\tau)\right\} \geq k \Big| (u_n(\tau))^{-1} X_0 = x\right)$$
$$\times Q_n(dx).$$

On the other hand, we have that

$$\left\{\sharp\left\{1 \leq j \leq q_n : X_0 \prod_{i=1}^j A_i + \Delta_j > u_n(\tau)\right\} \geq k\right\}$$
$$\subset \left\{\sharp\left\{1 \leq j \leq q_n : \left\{\left\{X_0 \prod_{i=1}^j A_i > u_n(\tau)(1 - \delta_n^{-1})\right\}\right.\right.\right.$$
$$\left.\left.\left. \cup \{\Delta_j > \delta_n^{-1} u_n(\tau)\}\right\} \geq k\right\}\right.$$
$$\subset \left\{\sharp\left\{1 \leq j \leq q_n : X_0 \prod_{i=1}^j A_i > u_n(\tau)(1 - \delta_n^{-1})\right\}\right\}$$

$$+ \sharp\{1 \leq j \leq q_n : \Delta_j > \delta_n^{-1} u_n(\tau)\} \geq k\Big\}$$

$$\subset \bigcup_{l=0}^{q_n} \{\sharp\{1 \leq j \leq q_n : \Delta_j > \delta_n^{-1} u_n(\tau)\} = l\}$$

$$\cap \{\sharp\{1 \leq j \leq q_n : \Delta_j > \delta_n^{-1} u_n(\tau)\} \geq (k-l) \vee 0\}.$$

Then

$$\int_1^\infty P\left(\sharp\left\{1 \leq j \leq q_n : X_0 \prod_{i=1}^j A_i + \Delta_j > u_n(\tau)\right\} \geq k \bigg| (u_n(\tau))^{-1} X_0 = x\right)$$

$$\times Q_n(dx)$$

$$\leq \int_1^\infty P\left(\sharp\left\{1 \leq j \leq q_n : X_0 \prod_{i=1}^j A_i > u_n(\tau)(1-\delta_n^{-1})\right\} \geq k \bigg|$$

$$(u_n(\tau))^{-1} X_0 = x\right) Q_n(dx)$$

$$+ \int_1^\infty P(\sharp\{1 \leq j \leq q_n : \{\Delta_j > \delta_n^{-1} u_n(\tau)\}\} > 0) Q_n(dx).$$

Note that $\Delta_j \leq X_j$ for $j \geq 1$ and, therefore,

$$\int_1^\infty P(\sharp\{1 \leq j \leq q_n : \{\Delta_j > \delta_n u_n(\tau)\}\} > 0) Q_n(dx)$$

$$\leq P(M_{q_n} > \delta_n^{-1} u_n(\tau)) \leq K \frac{q_{\lfloor n\delta_n^{-\kappa} \rfloor}}{n\delta_n^{-\kappa}}$$

if $q_n \to \infty$ and $n\delta_n^{-\kappa} \to \infty$ as $n \to \infty$. Moreover, by a change of variable, we have that

$$\int_1^\infty P\left(\sharp\left\{1 \leq j \leq q_n : X_0 \prod_{i=1}^j A_i > u_n(\tau)(1-\delta_n^{-1})\right\} \geq k \bigg| (u_n(\tau))^{-1} X_0 = x\right)$$

$$\times Q_n(dx)$$

$$= \frac{(1+o(1))}{(1-\delta_n^{-1})^\kappa} \int_{(1-\delta_n^{-1})^{-1}}^\infty P\left(\sharp\left\{1 \leq j \leq q_n : \prod_{i=1}^j A_i > x^{-1}\right\} \geq k\right) Q_n(dx).$$

Since the density function of $Q_n$ is uniformly bounded in a neighborhood of 1, we deduce that $\int_1^{(1-\delta_n^{-1})^{-1}} Q_n(dx) \leq K\delta_n^{-1}$ and it follows that

$$|B_k(q_n, (\Delta_j)_{j=1,\ldots,q_n}, Q_n) - B_k(q_n, (0)_{j=1,\ldots,q_n}, Q_n)| \leq K\left(\frac{q_{\lfloor n\delta_n^{-\kappa} \rfloor}}{n\delta_n^{-\kappa}} + \delta_n^{-1}\right).$$

(ii) Let $\varphi = EA_1^\xi < 1$ for $\xi \in (0, \kappa)$. We have that

$$\int_1^\infty P\left(\sharp\left\{j \geq 1 : \prod_{i=1}^j A_i > x^{-1}\right\} \geq k\right) Q_n(dx)$$

$$= \int_1^\infty P\left(\sharp\left\{1 \leq j \leq q_n : \prod_{i=1}^j A_i > x^{-1}\right\} + \left\{j > q_n : \prod_{i=1}^j A_i > x^{-1}\right\} \geq k\right)$$
$$\times Q_n(dx)$$

$$\leq \int_1^\infty P\left(\sharp\left\{1 \leq j \leq q_n : \prod_{i=1}^j A_i > x^{-1}\right\} \geq k\right) Q_n(dx)$$

$$+ \int_1^\infty P\left(\sharp\left\{j > q_n : \prod_{i=1}^j A_i > x^{-1}\right\} > 0\right) Q_n(dx).$$

It follows that

$$|B_k(q_n, (0)_{j=1,\ldots,q_n}, Q_n) - B_k(\infty, (0)_{j=1,\ldots,\infty}, Q_n)|$$

$$\leq \int_1^\infty P\left(\sharp\left\{j > q_n : \prod_{i=1}^j A_i > x^{-1}\right\} > 0\right) Q_n(dx)$$

$$\leq \int_1^{x_n} P\left(\bigvee_{j=q_n+1}^\infty \prod_{i=1}^j A_i > x^{-1}\right) Q_n(dx) + Q_n(x_n, \infty)$$

$$\leq \sum_{j=q_n+1}^\infty \varphi^j x_n^\xi + Q_n(x_n, \infty) \leq \frac{\varphi^{q_n} x_n^\xi}{1-\varphi} + \frac{K}{x_n^{\kappa-\epsilon}}$$

by Chebyshev's inequality and Potter's bounds.

(iii) Let $f_k(x) = P(\sharp\{\{j \geq 1 : \prod_{i=1}^j A_i > x^{-1}\} \geq k)$. Since the distribution of $A_1$ is absolutely continuous, $f_k$ is differentiable. Then we have

$$B_k(\infty, (0)_{j=1,\ldots,\infty}, Q_n) - B_k(\infty, (0)_{j=1,\ldots,\infty}, Q)$$

$$= \int_1^\infty f_k(x)(Q_n(dx) - Q(dx)) = \int_1^\infty f_k^{(1)}(x)(Q_n(x,\infty) - Q(x,\infty))\, dx,$$

where $f_k^{(1)}$ is the first derivative of $f_k$. But, by equation (3.2), $\sup_{x\geq 1}|(Q_n(x,\infty) \times Q^{-1}(x,\infty) - 1)| \leq Kn^{-\rho/\kappa}$ and we deduce that

$$|B_k(\infty, (0)_{j=1,\ldots,\infty}, Q_n) - B_k(\infty, (0)_{j=1,\ldots,\infty}, Q)| \leq Kn^{-\rho/\kappa}.$$

Putting the inequalities together yields

$$\text{II}b + \sum_{k=1}^{l+1} \text{II}e_k \leq K\left(\frac{q_{\lfloor n\delta_n^{-\kappa}\rfloor}}{n\delta_n^{-\kappa}} + \delta_n^{-1} + \varphi^{q_n} x_n^\xi + x_n^{-(\kappa-\epsilon)} + n^{-\rho/\kappa}\right)$$

and the result follows. □

We now define the big blocks $I_j^\triangle$ and the small blocks $I_j^*$ by

$$I_j^\triangle = \{(j-1)r_n + 1, \ldots, jr_n - l_n\},$$
$$I_j^* = \{jr_n - l_n + 1, \ldots, jr_n\}, \qquad j = 1, \ldots, k_n.$$

Let us introduce the following r.v.s associated with the big and small blocks:

$$N_{r_n,j}^{(\tau),\triangle} = \sum_{i \in I_j^\triangle} 1_{\{X_i > u_{r_n}(\tau)\}},$$

$$N_{r_n,j}^{(\tau),*} = \sum_{i \in I_j^*} 1_{\{X_i > u_{r_n}(\tau)\}}, \qquad j = 1, \ldots, k_n,$$

$$p_n^{(\tau),\triangle}(i) = \frac{1}{k_n} \sum_{j=1}^{k_n} 1_{\{N_{r_n,j}^{(\tau),\triangle} = i\}},$$

$$p_n^{(\tau),*}(i) = \frac{1}{k_n} \sum_{j=1}^{k_n} (1_{\{N_{r_n,j}^{(\tau),\triangle} = i - N_{r_n,j}^{(\tau),*}, N_{r_n,j}^{(\tau),*} > 0\}} - 1_{\{N_{r_n,j}^{(\tau),\triangle} = i, N_{r_n,j}^{(\tau),*} > 0\}}),$$

$$\bar{p}_n^{(\tau),\triangle} = \frac{1}{k_n} \sum_{j=1}^{k_n} N_{r_n,j}^{(\tau),\triangle},$$

$$\bar{p}_n^{(\tau),*} = \frac{1}{k_n} \sum_{j=1}^{k_n} N_{r_n,j}^{(\tau),*}.$$

It is easily seen that $p_n^{(\tau)}(i) = p_n^{(\tau),\triangle}(i) + p_n^{(\tau),*}(i)$ and $\bar{p}_n^{(\tau)} = \bar{p}_n^{(\tau),\triangle} + \bar{p}_n^{(\tau),*}$.

To prove Proposition 4.1, we will need the three following lemmas. The first lemma can be derived from Lemma 1 in [7].

LEMMA 6.3. *Let $p_1$, $p_2$, $p_3$ be positive numbers such that $p_1^{-1} + p_2^{-1} + p_3^{-1} = 1$. Suppose that $Y$ and $Z$ are random variables measurable with respect to the $\sigma$-algebra $\mathcal{F}_{1,m}^{(\tau_1,\ldots,\tau_r)}$, $\mathcal{F}_{m+l,n}^{(\tau_1,\ldots,\tau_r)}$ respectively $(1 \leq m \leq n-l)$ and assume further that $\|Y\|_{p_1} = (E|Y|^{p_1})^{1/p_1} < \infty$, $\|Z\|_{p_2} = (E|Z|^{p_2})^{1/p_2} < \infty$. Then*

$$|\operatorname{Cov}(Y,Z)| \leq 10(\alpha_{n,l}(\tau_1, \ldots, \tau_r))^{1/p_3} \|Y\|_{p_1} \|Z\|_{p_2}.$$

LEMMA 6.4. *Suppose that (C0) holds. Let $(r_n)$ be a sequence such that $r_n \to \infty$ and $r_n = o(n)$. Then $p_n^{(\tau)}(i) \xrightarrow{P} p^{(\tau)}(i)$ and $\bar{p}_n^{(\tau)} \xrightarrow{P} \tau$.*

PROOF. Since $r_n \to \infty$, $\Delta(u_{r_n}(\tau))$ holds and there exists a sequence $(l_n)$ such that $l_n \to \infty$, $l_n = o(r_n)$ and $\alpha_{r_n,l_n}(\tau) \to 0$. Let $\varepsilon > 0$. By Chebyshev's inequality,

$$P(|p_n^{(\tau),*}(i)| > \varepsilon)$$
$$\leq \varepsilon^{-1}(P(N_{r_n,1}^{(\tau),\triangle} = i - N_{r_n,1}^{(\tau),*}, N_{r_n,1}^{(\tau),*} > 0) + P(N_{r_n,1}^{(\tau),\triangle} = i, N_{r_n,1}^{(\tau),*} > 0))$$
$$\leq 2\varepsilon^{-1} P(N_{r_n,1}^{(\tau),*} > 0) \leq 2\varepsilon^{-1} P\left(\bigcup_{i \in I_1^*} \{X_i > u_{r_n}(\tau)\}\right) \leq 2\varepsilon^{-1} \tau l_n/r_n \to 0$$

and

$$P(|\bar{p}_n^{(\tau),*}| > \varepsilon) \leq \varepsilon^{-1} E(N_{r_n,1}^{(\tau),*}) \leq \varepsilon^{-1} \tau l_n/r_n \to 0.$$

Hence, $p_n^{(\tau),*}(i) \xrightarrow{P} 0$ and $\bar{p}_n^{(\tau),*} \xrightarrow{P} 0$. Now let us show that $p_n^{(\tau),\triangle}(i) \xrightarrow{P} p^{(\tau)}(i)$ and $\bar{p}_n^{(\tau),\triangle} \xrightarrow{P} \tau$. Since $\lim_{n\to\infty} P(N_{r_n,1}^{(\tau),*} = i) = 0$ and $\lim_{n\to\infty} E(N_{r_n,1}^{(\tau),*}) = 0$, we deduce by condition (C0.b) that $\lim_{n\to\infty} P(N_{r_n,1}^{(\tau),\triangle} = i) = p^{(\tau)}(i)$ and $\lim_{n\to\infty} E(N_{r_n,1}^{(\tau),\triangle}) = \tau$. Therefore, it suffices to show that

$$p_n^{(\tau),\triangle}(i) - P(N_{r_n,1}^{(\tau),\triangle} = i) \xrightarrow{P} 0 \quad \text{and} \quad \bar{p}_n^{(\tau),\triangle} - E(N_{r_n,1}^{(\tau),\triangle}) \xrightarrow{P} 0.$$

We have

$$P(|p_n^{(\tau),\triangle}(i) - P(N_{r_n,1}^{(\tau),\triangle} = i)| > \varepsilon)$$
$$\leq \varepsilon^{-2} E(p_n^{(\tau),\triangle}(i) - P(N_{r_n,1}^{(\tau),\triangle} = i))^2$$
$$\leq 2(k_n\varepsilon)^{-2} \sum_{1 \leq j \leq l \leq k_n} |\text{Cov}(1_{\{N_{r_n,l}^{(\tau),\triangle}=i\}}, 1_{\{N_{r_n,j}^{(\tau),\triangle}=i\}})|.$$

By using Lemma 6.3 with $p_1 = \infty$, $p_2 = \infty$, $p_3 = 1$, we get

$$P(|p_n^{(\tau),\triangle}(i) - P(N_{r_n,1}^{(\tau),\triangle} = i)| > \varepsilon)$$
$$\leq K(k_n\varepsilon)^{-2}\left(k_n + \sum_{j=1}^{k_n-1}(k_n-j)\alpha_{r_n,l_n+(j-1)r_n}(\tau)\right)\|1_{\{N_{r_n,1}^{(\tau),\triangle}=i\}}\|_\infty^2$$
$$\leq K\varepsilon^{-2}(k_n^{-1} + \alpha_{r_n,l_n}(\tau)) \to 0.$$

In the same way, by using Lemma 6.3 with $p_1 = \rho$, $p_2 = \rho$, $p_3 = \rho/(\rho-2)$, we get

$$P(|\bar{p}_n^{(\tau),\triangle} - E(N_{r_n,1}^{(\tau),\triangle})| > \varepsilon)$$
$$\leq \varepsilon^{-2} E(\bar{p}_n^{(\tau),\triangle} - E(N_{r_n,1}^{(\tau),\triangle}))^2 \leq 2(k_n\varepsilon)^{-2} \sum_{1 \leq j \leq l \leq k_n} |\text{Cov}(N_{r_n,j}^{(\tau),\triangle}, N_{r_n,l}^{(\tau),\triangle})|$$

$$\leq K(k_n \varepsilon)^{-2} \left( k_n + \sum_{j=1}^{k_n-1} (k_n - j)(\alpha_{r_n, l_n + (j-1)r_n}(\tau))^{1-2\rho^{-1}} \right) \|N_{r_n,1}^{(\tau), \triangle}\|_\rho^2$$

$$\leq K\varepsilon^{-2}(k_n^{-1} + (\alpha_{r_n, l_n}(\tau))^{1-2\rho^{-1}}) \|N_{r_n,1}^{(\tau)}\|_\rho^2.$$

Observe that $\sup_{n \geq 1} E(N_{r_n,1}^{(\tau)})^\rho < \infty$ by condition (C0.b) and $1 - 2\rho^{-1} > 0$ to conclude. $\square$

LEMMA 6.5. *Suppose that (C0) holds. Let $(r_n)$ be a sequence such that $r_n \to \infty$ and $r_n = o(n)$. Then*

$$(p_n^{(\cdot)}(0), \ldots, p_n^{(\cdot)}(m), \bar{p}_n^{(\cdot)}) \Rightarrow (p^{(\cdot)}(0), \ldots, p^{(\cdot)}(m), (\cdot)) \qquad in~ D^{m+2}.$$

PROOF. Let us first recall that convergence in $D^{m+2}$ is equivalent to convergence in $D_{\sigma, \phi}^{m+2}$ for all choice of positive $\sigma$ and $\phi$, $0 < \sigma < \phi < \infty$. Moreover, since $(p^{(\cdot)}(0), \ldots, p^{(\cdot)}(m), (\cdot))$ is a deterministic element of $D_{\sigma, \phi}^{m+2}$, we only need to prove that $p_n^{(\cdot)}(i) \Rightarrow p^{(\cdot)}(i)$ in $D_{\sigma, \phi}^1$, $i = 0, \ldots, m$, and $\bar{p}_n^{(\cdot)} \Rightarrow (\cdot)$ in $D_{\sigma, \phi}^1$. By Theorem 13.1 in [3], it suffices to prove that the finite-dimensional distributions converge and that a tightness criterion holds. It is easily seen that the first condition is satisfied by using Lemma 6.4. We only need to check that the $(p_n^{(\cdot)}(i))_{n \geq 1}$, $i = 0, \ldots, m$, and $(\bar{p}_n^{(\cdot)})_{n \geq 1}$ are tight in $D_{\sigma, \phi}^1$. Following Section 12 in [3], we call a set $\{\tau_i\}$ a $\delta$-sparse if it satisfies $\sigma = \tau_0 < \cdots < \tau_w = \phi$ and $\min_{1 \leq i \leq w}(\tau_i - \tau_{i-1}) \geq \delta$, and we define for $q \in D_{\sigma, \phi}^1$

$$w'(q, \delta) = \inf_{\{t_i\}} \max_{1 \leq i \leq w} \sup_{s, t \in (\tau_{i-1}, \tau_i]} |q(s) - q(t)|.$$

By using Theorem 13.2 in [3] and its corollary, $p_n^{(\cdot)}(i)$ is tight in $D_{\sigma, \phi}^1$ if and only if the two following conditions hold:

(i) for each $\tau$ in a set that is dense in $[\sigma, \phi]$ and contains $\sigma$,

$$\lim_{a \to \infty} \limsup_n P(p_n^{(\tau)}(i) > a) = 0,$$

(ii) for each $\varepsilon > 0$, $\lim_{\delta \to 0} \limsup_n P(w'(p_n^{(\cdot)}(i), \delta) > \varepsilon) = 0$.

Condition (i) is satisfied since $p_n^{(\tau)}(i) \xrightarrow{P} p^{(\tau)}(i) < 1$ for each $\tau \in [\sigma, \phi]$ (by Lemma 6.4). Let us now consider condition (ii). Let $\delta < \phi - \sigma$ and define $M_\delta = \lfloor (\phi - \sigma)\delta^{-1} \rfloor + 1$, $\tau_l^\delta = \sigma + l\delta$ for $0 \leq l < M_\delta$ and $\tau_{M_\delta}^\delta = \phi$. Note that $\tau \mapsto \sum_{j=0}^i p_n^{(\tau)}(j)$ is a nonincreasing function, and then

$$\sup_{\tau, \tau' \in (\tau_{l-1}^\delta, \tau_l^\delta]} \left| \sum_{j=0}^i (p_n^{(\tau)}(j) - p_n^{(\tau')}(j)) \right| \leq \sum_{j=0}^i (p_n^{(\tau_{l-1}^\delta)}(j) - p_n^{(\tau_l^\delta)}(j)).$$

It follows that
$$w'\left(\sum_{j=0}^{i} p_n^{(\cdot)}(j), \delta\right) \leq \max_{1 \leq l \leq M_\delta} \sum_{j=0}^{i}(p_n^{(\tau_{l-1}^\delta)}(j) - p_n^{(\tau_l^\delta)}(j)).$$

If $i \geq 1$, we have
$$P(w'(p_n^{(\cdot)}(i), \delta) > \varepsilon)$$
$$\leq P\left(w'\left(\sum_{j=0}^{i} p_n^{(\cdot)}(j), \delta\right) > \frac{\varepsilon}{2}\right) + P\left(w'\left(\sum_{j=0}^{i-1} p_n^{(\cdot)}(j), \delta\right) > \frac{\varepsilon}{2}\right)$$
$$\leq P\left(\max_{1 \leq l \leq M_\delta} \sum_{j=0}^{i}(p_n^{(\tau_{l-1}^\delta)}(j) - p_n^{(\tau_l^\delta)}(j)) > \frac{\varepsilon}{2}\right)$$
$$+ P\left(\max_{1 \leq l \leq M_\delta} \sum_{j=0}^{i-1}(p_n^{(\tau_{l-1}^\delta)}(j) - p_n^{(\tau_l^\delta)}(j)) > \frac{\varepsilon}{2}\right).$$

If $i = 0$, we have
$$P(w'(p_n^{(\cdot)}(i), \delta) > \varepsilon) \leq P\left(\max_{1 \leq l \leq M_\delta}(p_n^{(\tau_{l-1}^\delta)}(0) - p_n^{(\tau_l^\delta)}(0)) > \varepsilon\right).$$

By using Lemma 6.4, we get
$$\max_{1 \leq l \leq M_\delta} \sum_{j=0}^{i}(p_n^{(\tau_{l-1}^\delta)}(j) - p_n^{(\tau_l^\delta)}(j)) \xrightarrow{P} \max_{1 \leq l \leq M_\delta} \sum_{j=0}^{i}(p^{(\tau_{l-1}^\delta)}(j) - p^{(\tau_l^\delta)}(j)),$$

which is less than $\varepsilon/2$ for small $\delta$, since $\tau \mapsto \sum_{j=0}^{i} h_j(\tau)$ is a continuous and bounded function on $[\sigma, \phi]$. Thus, we deduce that
$$\lim_{\delta \to 0} \limsup_n P(w'(p_n^{(\cdot)}(i), \delta) > \varepsilon) = 0.$$

Condition (ii) is satisfied and $p_n^{(\cdot)}(i)$ is tight in $D_{\sigma,\phi}^1$.

Now note that $\tau \mapsto \bar{p}_n^{(\tau)}$ is a nondecreasing function and $\partial \bar{p}^{(\tau)}/\partial \tau = 1$. The arguments for $\bar{p}_n^{(\cdot)}$ run similarly. We conclude that $(p_n^{(\cdot)}(0), \ldots, p_n^{(\cdot)}(m), \bar{p}_n^{(\cdot)})$ weakly converges in $D_{\sigma,\phi}^{m+2}$, and then in $D^{m+2}$. □

PROOF OF PROPOSITION 4.1. The generalized inverse of $\bar{p}_n^{(\cdot)}$ is given by
$$\bar{p}_n^{(\bar{\tau}), \leftarrow} = \inf\left\{\tau \geq 0 : \sum_{i=1}^{r_n k_n} 1_{\{X_i > F^\leftarrow(1 - \tau/r_n)\}} \geq k_n \bar{\tau}\right\} = r_n \bar{F}(X_{k_n r_n - \lfloor k_n \bar{\tau} \rfloor : k_n r_n})$$

since $F^\leftarrow(F(X_{k_n r_n - \lfloor k_n \bar{\tau} \rfloor : k_n r_n})) = X_{k_n r_n - \lfloor k_n \bar{\tau} \rfloor : k_n r_n}$. It is a càglàd function on $[\sigma, \phi]$. Note that for $\bar{\tau} \in [\sigma, \phi]$ and $n$ such that $\lfloor k_n \bar{\tau} \rfloor \leq k_n r_n$,
$$\hat{p}_n^{(\bar{\tau})}(m) = p_n^{(\bar{p}_n^{(\bar{\tau}), \leftarrow})}(m), \qquad m \geq 0.$$

Let $D_{\uparrow,\sigma,\phi}$ (resp. $D^{\sigma,\phi}_{\uparrow,\sigma,\phi}, C_{\uparrow,\sigma,\phi}, C^{\sigma,\phi}_{\uparrow,\sigma,\phi}$) be the space of nondecreasing functions from $[\sigma,\phi]$ to $R$ (resp. nondecreasing functions from $[\sigma,\phi]$ to $[\sigma,\phi]$, continuous nondecreasing functions from $[\sigma,\phi]$ to $R$, continuous nondecreasing functions from $[\sigma,\phi]$ to $[\sigma,\phi]$).

Let us introduce the map $\Upsilon$ from $D_{\uparrow,\sigma,\phi}$ to $D^{\sigma,\phi}_{\uparrow,\sigma,\phi}$ taking $h$ into $\max(\sigma, \min(h^{\leftarrow},\phi))$. It is continuous at $C^{\sigma,\phi}_{\uparrow,\sigma,\phi}$. Let us denote by $\bar{p}^{(\cdot),\leftarrow}_{n,b}$ the function $\Upsilon(\bar{p}^{(\cdot)}_n)$. By Lemma 6.5 and the continuous mapping theorem (CMT), it follows that $\bar{p}^{(\cdot),\leftarrow}_{n,b} \Rightarrow \Upsilon((\cdot)) = (\cdot)$ in $D^{\sigma,\phi}_{\uparrow,\sigma,\phi}$.

Moreover, the composition map from $D^{m+1}_{\sigma,\phi} \times D^{\sigma,\phi}_{\uparrow,\sigma,\phi}$ to $D^{m+1}_{\sigma,\phi}$ taking $(g,h)$ into $g \circ h$ is continuous at $(g,h) \in C^{m+1}_{\sigma,\phi} \times C^{\sigma,\phi}_{\uparrow,\sigma,\phi}$ (see, e.g., [2], page 145). It follows by the CMT that

$$(p_n^{(\bar{p}^{(\cdot),\leftarrow}_{n,b})}(0),\ldots,p_n^{(\bar{p}^{(\cdot),\leftarrow}_{n,b})}(m)) \Rightarrow (p^{(\cdot)}(0),\ldots,p^{(\cdot)}(m))$$

in $D^{m+1}_{\sigma,\phi}$. Now we have

$$\sup_{\tau \in [\sigma,\phi]} |p_n^{(\bar{p}^{(\tau),\leftarrow}_n)}(j) - p_n^{(\bar{p}^{(\tau),\leftarrow}_{n,b})}(j)|$$
$$\leq \sup_{\tau,\bar{\tau} \in [\bar{p}^{(\sigma),\leftarrow}_n,\sigma]} |p_n^{(\tau)}(j) - p_n^{(\bar{\tau})}(j)| 1_{\{\bar{p}^{(\sigma),\leftarrow}_n < \sigma\}}$$
$$\vee \sup_{\tau,\bar{\tau} \in [\phi,\bar{p}^{(\phi),\leftarrow}_n]} |p_n^{(\tau)}(j) - p_n^{(\bar{\tau})}(j)| 1_{\{\bar{p}^{(\phi),\leftarrow}_n > \phi\}}.$$

Since the weak limit of $(p_n^{(\cdot)}(j))_{n\geq 1}$ is continuous at $\sigma$ and $\phi$, $\bar{p}^{(\sigma),\leftarrow}_n \xrightarrow{P} \sigma$ and $\bar{p}^{(\phi),\leftarrow}_n \xrightarrow{P} \phi$, we deduce that

$$\sup_{\tau \in [\sigma,\phi]} |p_n^{(\bar{p}^{(\bar{\tau}),\leftarrow}_n)}(j) - p_n^{(\bar{p}^{(\tau),\leftarrow}_{n,b})}(j)| \xrightarrow{P} 0,$$

or, equivalently, $p_n^{(\bar{p}^{(\cdot),\leftarrow}_n)}(j) - p_n^{(\bar{p}^{(\cdot),\leftarrow}_{n,b})}(j) \Rightarrow 0$ in $D^1_{\sigma,\phi}$. Finally, we get

$$(\hat{p}^{(\cdot)}_n(0),\ldots,\hat{p}^{(\cdot)}_n(m)) \Rightarrow (p^{(\cdot)}(0),\ldots,p^{(\cdot)}(m)) \quad \text{in } D^{m+1}_{\sigma,\phi}.$$

To prove weak convergence of $(\hat{\pi}^{(\cdot)}_n(1),\ldots,\hat{\pi}^{(\cdot)}_n(m))$ in $D^m_{\sigma,\phi}$, we proceed by induction. First note that by Lemma 6.5 $\lim_{n\to\infty} P([\hat{p}^{(\phi)}_n(0),\hat{p}^{(\sigma)}_n(0)] \in (0,1)) = 1$. We deduce by the CMT that

$$\chi^{(\cdot)}_n(1) = -\frac{\hat{p}^{(\cdot)}_n(1)}{\ln(\hat{p}^{(\cdot)}_n(0))\hat{p}^{(\cdot)}_n(0)} \Rightarrow -\frac{p^{(\cdot)}(1)}{\ln(p^{(\cdot)}(0))p^{(\cdot)}(0)} = \pi(1)$$

in $D^1_{\sigma,\phi}$,

$$\hat{\pi}_n^{(\cdot)}(1) = \max(0, \min(\chi_n^{(\cdot)}(1), 1)) \Rightarrow \pi(1)$$

in $D^1_{\sigma,\phi}$ and

$$(\hat{p}_n^{(\cdot)}(0), \hat{p}_n^{(\cdot)}(1), \hat{\pi}_n^{(\cdot)}(1)) \Rightarrow (p^{(\cdot)}(0), p^{(\cdot)}(1), \pi(1))$$

in $D^3_{\sigma,\phi}$. Now assume that we have already shown that

$$(\hat{p}_n^{(\cdot)}(0), \ldots, \hat{p}_n^{(\cdot)}(j), \hat{\pi}_n^{(\cdot)}(1), \ldots, \hat{\pi}_n^{(\cdot)}(j-1))$$
$$\Rightarrow (p^{(\cdot)}(0), \ldots, p^{(\cdot)}(j), \pi(1), \ldots, \pi(j-1))$$

in $D^{2j}_{\sigma,\phi}$. Let us define the maps $\Psi_j$ from $D^{2j}_{\sigma,\phi}$ to $D^1_{\sigma,\phi}$ taking $f(\cdot) = (f_i(\cdot))_{i=1,\ldots,2j}$ into

$$\Psi_j(f(\cdot)) = -\frac{(f_{j+1}(\cdot) + j^{-1}\ln(f_1(\cdot))\sum_{i=1}^{j-1} i f_{i+j+1}(\cdot) f_{j-i+1}(\cdot)}{\ln(f_1(\cdot)) f_1(\cdot)}.$$

Note that

$$\chi_n^{(\cdot)}(j) = \Psi_j(\hat{p}_n^{(\cdot)}(0), \ldots, \hat{p}_n^{(\cdot)}(j), \hat{\pi}_n^{(\cdot)}(1), \ldots, \hat{\pi}_n^{(\cdot)}(j-1))$$

and that $\Psi_j$ is continuous on the space of continuous functions from $[\sigma, \phi]$ to $(0,1) \times R^{2j-1}$. It follows by the CMT that $\chi_n^{(\cdot)}(j) \Rightarrow \pi(j)$ in $D^1_{\sigma,\phi}$. Let us recall that

$$\hat{\pi}_n^{(\cdot)}(j) = \max\left(0, \min\left(\chi_n^{(\cdot)}(j), 1 - \sum_{i=1}^{j-1} \hat{\pi}_n^{(\cdot)}(i)\right)\right).$$

We conclude by the CMT that $\hat{\pi}_n^{(\cdot)}(j) \Rightarrow \pi(j)$ in $D^1_{\sigma,\phi}$ and

$$(\hat{p}_n^{(\cdot)}(0), \ldots, \hat{p}_n^{(\cdot)}(j+1), \hat{\pi}_n^{(\cdot)}(1), \ldots, \hat{\pi}_n^{(\cdot)}(j))$$
$$\Rightarrow (p^{(\cdot)}(0), \ldots, p^{(\cdot)}(j+1), \pi(1), \ldots, \pi(j))$$

in $D^{2(j+1)}_{\sigma,\phi}$. The induction is established and

$$(\hat{\pi}_n^{(\cdot)}(1), \ldots, \hat{\pi}_n^{(\cdot)}(m)) \Rightarrow (\pi(1), \ldots, \pi(m))$$

in $D^m_{\sigma,\phi}$. Finally, by using again the CMT, we deduce that

$$(\hat{\theta}_{1,n}^{(\cdot)}, \hat{\theta}_{2,n}^{(\cdot)}(m), \hat{\theta}_{3,n}^{(\cdot)}(m)) \Rightarrow (\theta, \theta_2(m), \theta_3^{(\cdot)}(m))$$

in $D^3_{\sigma,\phi}$, $\widehat{\bar{\pi}}_n(m) \xrightarrow{P} \pi(m)$, $m \geq 1$, and $\widehat{\bar{\theta}}_{1,n} \xrightarrow{P} \theta$. □

Let us now define

$$e_{i,n}^{\triangle}(\tau) = \sqrt{k_n}(p_n^{(\tau),\triangle}(i) - P(N_{r_n,j}^{(\tau),\triangle} = i)),$$

$$e_{i,n}^{*}(\tau) = \sqrt{k_n}(1_{\{N_{r_n,j}^{(\tau),\triangle} = i - N_{r_n,j}^{(\tau),*}, N_{r_n,j}^{(\tau),*} > 0\}} - P(N_{r_n,j}^{(\tau),\triangle} = i - N_{r_n,j}^{(\tau),*}, N_{r_n,j}^{(\tau),*} > 0))$$

$$- \sqrt{k_n}(1_{\{N_{r_n,j}^{(\tau),\triangle} = i, N_{r_n,j}^{(\tau),*} > 0\}} - P(N_{r_n,j}^{(\tau),\triangle} = i, N_{r_n,j}^{(\tau),*} > 0)),$$

$$\bar{e}_n^{\triangle}(\tau) = \sqrt{k_n}(\bar{p}_n^{(\tau),\triangle} - (r_n - l_n)P(X_i > u_{r_n}(\tau))),$$

$$\bar{e}_n^{*}(\tau) = \sqrt{k_n}(\bar{p}_n^{(\tau),*} - l_n P(X_i > u_{r_n}(\tau))),$$

$$E_{m,n}^{\triangle}(\tau) = (e_{0,n}^{\triangle}(\tau), \ldots, e_{m,n}^{\triangle}(\tau), \bar{e}_n^{\triangle}(\tau)),$$

$$E_{m,n}^{*}(\tau) = (e_{0,n}^{*}(\tau), \ldots, e_{m,n}^{*}(\tau), \bar{e}_n^{*}(\tau)).$$

We have $e_{j,n}(\cdot) = e_{j,n}^{\triangle}(\cdot) + e_{j,n}^{*}(\cdot)$ and $\bar{e}_n(\cdot) = \bar{e}_n^{\triangle}(\cdot) + \bar{e}_n^{*}(\cdot)$. The proof of Theorem 4.1 is now presented in a series of three lemmas.

LEMMA 6.6. *Suppose that (C2) holds. Let $\tau > 0$. Then $E_{m,n}^{*}(\tau) \xrightarrow{P} 0$.*

PROOF. By (C2.c), there exists a sequence $(l_n)$ satisfying $l_n = o(r_n^{2/r})$ and $\lim_{n \to \infty} n r_n^{-1} \alpha_{l_n} = 0$. We have that

$$E(e_{i,n}^{*}(\tau))^2$$

$$\leq 2k_n^{-1} E\left(\sum_{j=1}^{k_n}(1_{\{N_{r_n,j}^{(\tau),\triangle} = i - N_{r_n,j}^{(\tau),*}, N_{r_n,j}^{(\tau),*} > 0\}}\right.$$

$$\left. - P(N_{r_n,j}^{(\tau),\triangle} = i - N_{r_n,j}^{(\tau),*}, N_{r_n,j}^{(\tau),*} > 0))\right)^2$$

$$+ 2k_n^{-1} E\left(\sum_{j=1}^{k_n}(1_{\{N_{r_n,j}^{(\tau),\triangle} = i, N_{r_n,j}^{(\tau),*} > 0\}} - P(N_{r_n,j}^{(\tau),\triangle} = i, N_{r_n,j}^{(\tau),*} > 0))\right)^2$$

$$=: 2(I_1 + I_2).$$

Let $2 < v < r$. By using Lemma 6.3 with $p_1 = v$, $p_2 = v$, $p_3 = v/(v-2)$, we get

$$I_1 \leq 2k_n^{-1} \sum_{1 \leq j \leq l \leq k_n} \text{Cov}(1_{\{N_{r_n,j}^{(\tau),\triangle} = i - N_{r_n,j}^{(\tau),*}, N_{r_n,j}^{(\tau),*} > 0\}}, 1_{\{N_{r_n,l}^{(\tau),\triangle} = i - N_{r_n,l}^{(\tau),*}, N_{r_n,l}^{(\tau),*} > 0\}})$$

$$\leq K k_n^{-1}\left(k_n + \sum_{j=1}^{k_n-1}(k_n - j)(\alpha_{r_n,(j-1)r_n}(\tau))^{1-2v^{-1}}\right)$$

$$\times \|1_{\{N^{(\tau),\triangle}_{r_n,1}=i-N^{(\tau),*}_{r_n,1}, N^{(\tau),*}_{r_n,1}>0\}}\|_v^2$$

$$\leq K\left(1+\sum_{j=1}^{k_n-1}\alpha_{(j-1)r_n}^{1-2v^{-1}}\right)(P(N^{(\tau),*}_{r_n,1}>0))^{2/v}$$

$$\leq K\left(1+\sum_{j=0}^{\infty}\alpha_j^{1-2v^{-1}}\right)(l_n\bar{F}(u_{r_n}(\tau)))^{2/v} \leq K\left(\frac{l_n}{r_n}\right)^{2/v},$$

since $\sum_{j=0}^{\infty}\alpha_j^{1-2v^{-1}} < \infty$. Similarly, $I_2 \leq K(l_n/r_n)^{2/v}$. Therefore,

$$P(|e^*_{i,n}(\tau)|>\varepsilon) \leq \varepsilon^{-2}E(e^*_{i,n}(\tau))^2 \leq K(l_n/r_n)^{2/v} \to 0.$$

By using Lemma 6.3 with $p_1 = v$, $p_2 = v$, $p_3 = v/(v-2)$, we get

$$E(\bar{e}^*_n(\tau))^2 \leq \frac{K}{k_n}\left(k_n + \sum_{j=1}^{k_n-1}(k_n-j)(\alpha_{r_n,(r_n-l_n)+(j-1)r_n}(\tau))^{1-2v^{-1}}\right)$$

$$\times \|N^{(\tau),*}_{r_n,1} - l_n\bar{F}(u_{r_n}(\tau))\|_v^2$$

$$\leq K\left(1+\sum_{j=1}^{k_n-1}(\alpha_{r_n,(r_n-l_n)+(j-1)r_n}(\tau))^{1-2v^{-1}}\right)$$

$$\times \|N^{(\tau),*}_{r_n,1} - l_n\bar{F}(u_{r_n}(\tau))\|_v^2.$$

By Theorem 4.1 in [40] [equation (4.4)], we have

$$E|N^{(\tau),*}_{r_n,1} - l_n\bar{F}(u_{r_n}(\tau))|^v \leq Kl_n^{v/2}\|1_{\{X_1>u_{r_n}(\tau)\}} - \bar{F}(u_{r_n}(\tau))\|_r^v$$

$$\leq K\left(\frac{l_n}{r_n^{2/r}}\right)^{v/2} \to 0.$$

Putting the inequalities above together yields $E^*_{m,n}(\tau) \xrightarrow{P} 0$. □

LEMMA 6.7. *Suppose that (C1) and (C2) hold. Let $r \geq 1$ and $\tau_1 > \cdots > \tau_r > 0$. Then*

$$(E_{m,n}(\tau_1),\ldots,E_{m,n}(\tau_r)) \xrightarrow{d} (E_m(\tau_1),\ldots,E_m(\tau_r)).$$

PROOF. Since by Lemma 6.6 $E^*_{m,n}(\tau) \xrightarrow{P} 0$, we only prove that

$$(E^{\triangle}_{m,n}(\tau_1),\ldots,E^{\triangle}_{m,n}(\tau_r)) \xrightarrow{d} (E_m(\tau_1),\ldots,E_m(\tau_r)).$$

By applying the Cramer–Wold device, it suffices to prove that, for $\lambda_{h,j} \in R$, $h = 1,\ldots,r$ and $i = 0,\ldots,m+1$,

$$\sum_{h=1}^{r}\left(\sum_{i=0}^{m}\lambda_{h,i}e^{\triangle}_{i,n}(\tau_h) + \lambda_{h,m+1}\bar{e}^{\triangle}_n(\tau_h)\right) \xrightarrow{d} \sum_{h=1}^{r}\left(\sum_{i=0}^{m}\lambda_{h,i}e_i(\tau_h) + \lambda_{h,m+1}\bar{e}(\tau_h)\right).$$

Let

$$f_{j,n} = \sum_{h=1}^{r}\sum_{i=0}^{m}\lambda_{h,i}(1_{\{N_{r_n,j}^{(\tau_h),\triangle}=i\}} - P(N_{r_n,j}^{(\tau_h),\triangle}=i))$$

$$+ \sum_{h=1}^{r}\lambda_{h,m+1}(N_{r_n,j}^{(\tau_h),\triangle} - (r_n - l_n)P(X_1 > u_{r_n}(\tau_h))).$$

By using recursively Lemma 6.3 with $p_1 = \infty$, $p_2 = \infty$, $p_3 = 1$, we get

$$\left|E\exp\left\{-\frac{\mathbf{i}u}{\sqrt{k_n}}\sum_{j=1}^{k_n}f_{j,n}\right\} - \prod_{j=1}^{k_n}E\exp\left\{-\frac{\mathbf{i}u}{\sqrt{k_n}}f_{j,n}\right\}\right| \leq Kk_n\alpha_{r_n,l_n}(\tau_1,\ldots,\tau_r),$$

which tends to 0 by condition (C2.c). This implies that the $f_{j,n}$ can be considered as i.i.d. r.v.s. By condition (C0.b) and Minkowski's inequality, $\lim_{n\to\infty}E|f_{j,n}|^\rho < \infty$ where $\rho > 2$. Therefore,

$$\frac{\sum_{j=1}^{k_n}E|f_{j,n}|^\rho}{(\sum_{j=1}^{k_n}E(f_{j,n})^2)^{\rho/2}} = \frac{1}{k_n^{\rho/2-1}}\frac{E|f_{1,n}|^\rho}{(E(f_{1,n})^2)^{\rho/2}} \to 0$$

and Lyapounov's condition holds (see, e.g., [4], page 362). It follows that $(k_nE(f_{1,n})^2)^{-1/2}\sum_{i=1}^{k_n}f_{i,n}$ converges in distribution to a standard Gaussian random variable.

By Condition (C1), $(N_n^{(\tau_1)}(E), N_n^{(\tau_2)}(E)) \Rightarrow (N_E^{(\tau_1)}, N_E^{(\tau_2)})$ and the limiting second central moments of the r.v.s $1_{\{N_{r_n,1}^{(\tau_h),\triangle}=i\}}$ and $N_{r_n,1}^{(\tau_h),\triangle}$, $h=1,\ldots,r$, exist. Simple calculations yield the covariance functions given in Theorem 4.1. □

LEMMA 6.8. *Suppose that (C1) and (C2) hold. Then $(E_{m,n}(\cdot))_{n\geq 1}$ is tight in $D_{\sigma,\phi}^{m+2}$.*

PROOF. We use similar arguments as for the second part of the proof of Theorem 22.1 in [2]. The tightness criterion which is considered is the following (see Theorem 15.5 and Theorem 8.3 in [2]): $(E_{m,n}(\cdot))_{n\geq 1}$ is tight in $D_{\sigma,\phi}^{m+2}$ if:

(i) for each positive $\eta$, there exists an $a$ such that

$$P(|E_{m,n}(\phi)|_1 > a) \leq \eta, \qquad n \geq 1,$$

where $|E|_1 = \sum_{j=0}^{m+1}|E_j|$;

(ii) letting $\varepsilon > 0$ and $\eta > 0$, there exists $\delta > 0$ and an integer $n_0$ such that

$$P\left(\sup_{\tau_2\leq\tau_1\leq\tau_2+\delta}|E_{m,n}(\tau_1) - E_{m,n}(\tau_2)|_1 > \varepsilon\right) \leq \eta\delta, \qquad n \geq n_0,$$

for all $\tau_2 \in [\sigma,\phi]$.

CLUSTER SIZE DISTRIBUTION OF EXTREME VALUES 35Moreover, by Theorem 15.5 in [2], it follows that the weak limit of a subsequence $E_{m,n'}(\cdot)$ belongs a.s. to $C_{\sigma,\phi}^{m+2}$.

Condition (i) is satisfied since $E_{m,n}(\phi) \xrightarrow{d} E_m(\phi)$. Let us consider condition (ii). Note that

$$P\left(\sup_{\tau_2 \leq \tau_1 \leq \tau_2+\delta} |E_{m,n}(\tau_1) - E_{m,n}(\tau_2)|_1 > \varepsilon\right)$$

$$\leq \sum_{i=0}^{m} P\left(\sup_{\tau_2 \leq \tau_1 \leq \tau_2+\delta} |e_{i,n}(\tau_1) - e_{i,n}(\tau_2)| > \frac{\varepsilon}{m+1}\right)$$

$$+ P\left(\sup_{\tau_2 \leq \tau_1 \leq \tau_2+\delta} |\bar{e}_n(\tau_1) - \bar{e}_n(\tau_2)| > \frac{\varepsilon}{m+1}\right)$$

$$\leq 2\sum_{i=0}^{m} P\left(\sup_{\tau_2 \leq \tau_1 \leq \tau_2+\delta} \left|\sum_{j=0}^{i}(e_{j,n}(\tau_1) - e_{j,n}(\tau_2))\right| > \frac{\varepsilon}{2(m+1)}\right)$$

$$+ P\left(\sup_{\tau_2 \leq \tau_1 \leq \tau_2+\delta} |\bar{e}_n(\tau_1) - \bar{e}_n(\tau_2)| > \frac{\varepsilon}{m+1}\right)$$

and it suffices to check the tightness criterion for each $\sum_{j=0}^{i} e_{j,n}(\cdot)$, $i = 0, \ldots, m$ and for $\bar{e}_n(\cdot)$. Now we simply indicate the modifications to be made in the proof of Theorem 22.1 in [2] to establish that condition (ii) holds.

Let $2 < v < p < r \leq \infty$ and $\varepsilon > 0$. Assume that $\theta_d > v/(v-2)$ and $\theta_d \geq (p-1)r/(r-p)$.

(i) Let $\sigma \leq \tau_2 < \tau_1 \leq \phi$ and define

$$S_i(\tau_1, \tau_2; k_n) := \sqrt{k_n}\left(\sum_{j=0}^{i}(e_{j,n}(\tau_1) - e_{j,n}(\tau_2))\right).$$

By Theorem 4.1 in [40] [equation (4.3)], we have that

$E|S_i(\tau_1, \tau_2; k_n)|^p$

$\leq K(k_n^{p/2}(P(N_{r_n,1}^{(\tau_2)} \leq i < N_{r_n,1}^{(\tau_1)}))^{p/v} + k_n^{1+\varepsilon}(P(N_{r_n,1}^{(\tau_2)} \leq i < N_{r_n,1}^{(\tau_1)}))^{p/r})$

$\leq K(k_n^{p/2}(P(N_{r_n,1}^{(\tau_1)} - N_{r_n,1}^{(\tau_2)} > 1))^{p/v} + k_n^{1+\varepsilon}(P(N_{r_n,1}^{(\tau_1)} - N_{r_n,1}^{(\tau_2)} > 1))^{p/r})$

$\leq K(k_n^{p/2}(E(N_{r_n,1}^{(\tau_1)} - N_{r_n,1}^{(\tau_2)}))^{p/v} + k_n^{1+\varepsilon}(E(N_{r_n,1}^{(\tau_1)} - N_{r_n,1}^{(\tau_2)}))^{p/r})$

$\leq K(k_n^{p/2}(\tau_1 - \tau_2)^{p/v} + k_n^{1+\varepsilon}(\tau_1 - \tau_2)^{p/r}).$

Let $\eta = p/2 - (1+\varepsilon)$. If $0 < \epsilon < 1$ and $\epsilon/k_n^{\eta} \leq (\tau_1 - \tau_2)^{p(1/v - 1/r)}$, we get

$$E\left(\left|\sum_{j=0}^{i}(e_{j,n}(\tau_1) - e_{j,n}(\tau_2))\right|^p\right) \leq K\epsilon^{-1}(\tau_1 - \tau_2)^{p/v},$$

which replaces equation (22.15) of [2].

(ii) Let $\xi_{j,n} := (N_{r_n,j}^{(\tau_1)} - N_{r_n,j}^{(\tau_2)} - (EN_{r_n,j}^{(\tau_1)} - EN_{r_n,j}^{(\tau_2)}))$ and define

$$S(\tau_1, \tau_2; k_n) := \sqrt{k_n}(\bar{e}_n(\tau_1) - \bar{e}_n(\tau_2)) = \sum_{j=1}^{k_n} \xi_{j,n}.$$

By Theorem 4.1 in [40] [equation (4.3)], we have that

$$E|S(\tau_1, \tau_2; k_n)|^p \leq K(k_n^{p/2}\|\xi_{1,n}\|_v^p + k_n^{1+\varepsilon}\|\xi_{1,n}\|_r^p).$$

Now for $v > 2$,

$$|\xi_{1,n}|^v \leq 2^v((N_{r_n,1}^{(\tau_1)} - N_{r_n,1}^{(\tau_2)})^v + (EN_{r_n,1}^{(\tau_1)} - EN_{r_n,1}^{(\tau_2)})^v).$$

For large $n$ and for $\sigma \leq \tau_2 < \tau_1 \leq \phi$,

$$|\xi_{1,n}|^v \leq K((N_{r_n,1}^{(\tau_1)} - N_{r_n,1}^{(\tau_2)})^v + (\tau_1 - \tau_2)).$$

By condition (C2.a), we get $E(|\xi_{1,n}|^\lambda) \leq K(\tau_1 - \tau_2)$ for $2 \leq \lambda \leq r$ and we deduce that

$$E|S(\tau_1, \tau_2; k_n)|^p \leq K(k_n^{p/2}(\tau_1 - \tau_2)^{p/v} + k_n^{1+\varepsilon}(\tau_1 - \tau_2)^{p/r}).$$

Therefore, if $\epsilon < 1$ and $\epsilon/k_n^\eta \leq (\tau_1 - \tau_2)^{p(1/v - 1/r)}$, we have that

$$E(|\bar{e}_n(\tau_1) - \bar{e}_n(\tau_2)|^p) \leq K\epsilon^{-1}(\tau_1 - \tau_2)^{p/v},$$

which also replaces equation (22.15) of [2].

(iii) We replace equation (22.17) in [2] by

$$\left|\sum_{j=0}^{i}(e_{j,n}(\tau_1) - e_{j,n}(\tau_2))\right| \leq \left|\sum_{j=0}^{i}(e_{j,n}(\tau_2 + \delta) - e_{j,n}(\tau_2))\right| + \delta\sqrt{k_n},$$

$$|\bar{e}_n(\tau_1) - \bar{e}_n(\tau_2)| \leq |\bar{e}_n(\tau_2 + \delta) - \bar{e}_n(\tau_2)| + \delta\sqrt{k_n},$$

for $\tau_2 \leq \tau_1 \leq \tau_2 + \delta$, by using monotony arguments as in [2].

(iv) We need to replace (22.19) of [2] by

$$\left(\frac{\epsilon}{k_n^\eta}\right)^{rv/(p(r-v))} \leq p < \frac{\epsilon}{\sqrt{k_n}}$$

and to assume that

$$\frac{\eta r v}{p(r-v)} = \frac{rv}{(r-v)}\left(\frac{1}{2} - \frac{(1+\varepsilon)}{p}\right) > \frac{1}{2}.$$

Since $\theta_d$ has to be larger than $(p-1)r/(r-p)$ which is increasing in $p$ and $p > v$, we let $p = v(1+\varepsilon)$ and choose $v$ such that

$$\frac{rv}{(r-v)}\left(\frac{1}{2} - \frac{(1+\varepsilon)}{p}\right) = \frac{1}{2}(1+\varepsilon).$$

It follows that $v = (3+\varepsilon)r/(r+(1+\varepsilon))$. Then the inequalities $\theta_d > v/(v-2)$ and $\theta_d \geq (p-1)r/(r-p)$ become

$$\theta_d > \frac{3+\varepsilon}{1+\varepsilon}\frac{r}{r-2} \quad \text{and} \quad \theta_d \geq \frac{((2+\varepsilon)^2 - 2)r - (1+\varepsilon)}{r - (2+\varepsilon)(1+\varepsilon)},$$

which are satisfied if $\varepsilon < ((r-2) \wedge 1/2)/4$ and

$$\theta_d \geq \frac{3r}{r - 2(1+2\varepsilon)}.$$

Everything else remains the same as for the proof of Theorem 22.1 in [2]. Finally, choose $\mu = 4\varepsilon$. □

PROOF OF THEOREM 4.1. Weak convergence in $D^{m+2}$ of a stochastic process is equivalent to weak convergence of the restrictions of the stochastic process to any compact $[\sigma, \phi]$ with $0 < \sigma < \phi < \infty$ in $D_{\sigma,\phi}^{m+2}$. The convergence of the finite dimensional distributions of $E_{m,n}(\cdot)$ is established by Lemma 6.7 and the tightness of $(E_{m,n}(\cdot))_{n\geq 1}$ in $D_{\sigma,\phi}^{m+2}$ by Lemma 6.8. Weak convergence in $D_{\sigma,\phi}^{m+2}$ follows by Theorem 13.1 in [3]. By Theorem 15.5 in [2], we deduce that $E_m(\cdot) \in C^{m+2}$. □

PROOF OF THEOREM 4.2. Let

$$\tilde{e}_{j,n}(\cdot) := \sqrt{k_n}(p_n^{(\cdot)}(j) - p^{(\cdot)}(j)) = e_{j,n}(\cdot) + \sqrt{k_n}(P(N_{r_n,1}^{(\cdot)} = j) - p^{(\cdot)}(j)).$$

Since $\sup_{\tau\in[\sigma,\phi]} |\sqrt{k_n}(P(N_{r_n,1}^{(\tau)} = j) - p^{(\tau)}(j))| \to 0$ [by Condition (C3)], we deduce that $(\tilde{e}_{0,n}(\cdot), \ldots, \tilde{e}_{m,n}(\cdot)) \Rightarrow (e_0(\cdot), \ldots, e_m(\cdot))$ in $D_{\sigma,\phi}^{m+1}$. By using the function $\Upsilon$, the composition map, the same arguments as in the proof of Proposition 4.1 and Theorem 4.1, we deduce that

$$(\tilde{e}_{0,n}(\bar{p}_{n,b}^{(\cdot),\leftarrow}), \ldots, \tilde{e}_{m,n}(\bar{p}_{n,b}^{(\cdot),\leftarrow})) \Rightarrow (e_0(\cdot), \ldots, e_m(\cdot))$$

in $D_{\sigma,\phi}^{m+1}$. Now note that

$$\sup_{\tau\in[\sigma,\phi]} |\tilde{e}_{j,n}(\bar{p}_n^{(\cdot),\leftarrow}) - \tilde{e}_{j,n}(\bar{p}_{n,b}^{(\cdot),\leftarrow})|$$

$$\leq \sup_{\tau,\bar{\tau}\in[\bar{p}_n^{(\sigma),\leftarrow},\sigma]} |\tilde{e}_{j,n}(\tau) - \tilde{e}_{j,n}(\bar{\tau})| 1_{\{\bar{p}_n^{(\sigma),\leftarrow} < \sigma\}}$$

$$\vee \sup_{\tau,\bar{\tau}\in[\phi,\bar{p}_n^{(\phi),\leftarrow}]} |\tilde{e}_{j,n}(\tau) - \tilde{e}_{j,n}(\bar{\tau})| 1_{\{\bar{p}_n^{(\phi),\leftarrow} > \phi\}}.$$

Since the weak limit of $(\tilde{e}_{j,n}(\cdot))_{n\geq 1}$ is continuous at $\sigma$ and $\phi$, $\bar{p}_n^{(\sigma),\leftarrow} \xrightarrow{P} \sigma$ and $\bar{p}_n^{(\phi),\leftarrow} \xrightarrow{P} \phi$, it follows that $\sup_{\tau\in[\sigma,\phi]} |\tilde{e}_{j,n}(\bar{p}_n^{(\cdot),\leftarrow}) - \tilde{e}_{j,n}(\bar{p}_{n,b}^{(\cdot),\leftarrow})| \xrightarrow{P} 0$ and that $\tilde{e}_{j,n}(\bar{p}_n^{(\cdot),\leftarrow}) - \tilde{e}_{j,n}(\bar{p}_{n,b}^{(\cdot),\leftarrow}) \Rightarrow 0$ in $D_{\sigma,\phi}^1$. Let

$$\tilde{e}_n(\cdot) := \sqrt{k_n}(\bar{p}_n^{(\cdot)} - (\cdot)) = \bar{e}_n(\cdot) + \sqrt{k_n}(r_n\bar{F}(u_{r_n}(\cdot)) - (\cdot)).$$

By Condition (C3), $\sup_{\tau \in [\sigma, \phi]} \sqrt{k_n} |r_n \bar{F}(u_{r_n}(\tau)) - \tau| \to 0$. It follows by Theorem 4.1 that $\tilde{e}_n(\cdot) \Rightarrow \bar{e}(\cdot)$ in $D^1_{\sigma,\phi}$. Now by using Vervaat's lemma [42], we get

$$\sqrt{k_n}(\bar{p}_n^{(\cdot),\leftarrow} - (\cdot)) \Rightarrow -\bar{e}(\cdot) \qquad \text{in } D^1_{\sigma,\phi}.$$

We deduce from the differentiability of $p^{(\cdot)}(j)$ and the finite increments formula that

$$\sqrt{k_n}(p^{(\bar{p}_n^{(\cdot),\leftarrow})}(j) - p^{(\cdot)}(j)) \Rightarrow -h_j(\cdot)\bar{e}(\cdot)$$

in $D^1_{\sigma,\phi}$. Finally, we get

$$\hat{e}_{j,n}(\cdot) = (\tilde{e}_{j,n}(\bar{p}_n^{(\cdot),\leftarrow}) - \tilde{e}_{j,n}(\bar{p}_{n,b}^{(\cdot),\leftarrow})) + \tilde{e}_{j,n}(\bar{p}_{n,b}^{(\cdot),\leftarrow})$$
$$+ \sqrt{k_n}(p^{(\bar{p}_n^{(\cdot),\leftarrow})}(j) - p^{(\cdot)}(j))$$
$$\Rightarrow e_j(\cdot) - h_j(\cdot)\bar{e}(\cdot) = \hat{e}_j(\cdot)$$

in $D^1_{\sigma,\phi}$ and

$$(\hat{e}_{0,n}(\cdot), \ldots, \hat{e}_{m,n}(\cdot)) \Rightarrow (\hat{e}_0(\cdot), \ldots, \hat{e}_m(\cdot)) \qquad \text{in } D^{m+1}_{\sigma,\phi}. \qquad \square$$

PROOF OF COROLLARY 4.1.  We first recall that a map $T$ between topological vector spaces $B_i$, $i = 1, 2$, is called Hadamard differentiable tangentially to some subset $S \subset B_1$ at $x \in B_1$ if there exists a continuous linear map $T'(x)$ from $B_1$ to $B_2$ such that

$$\frac{T(x + t_n y_n) - T(x)}{t_n} \to T'(x) \cdot y$$

for all sequences $t_n \downarrow 0$ and $y_n \in B_1$ converging to $y \in S$. Note that the map $\Psi_j$ introduced in the proof of Proposition 4.1 is Hadamard differentiable tangentially to $C^{2j}_{\sigma,\phi}$ at $f \in C^{2j}_{\sigma,\phi}$ and that

$$\Psi'_j(f(\cdot)) \cdot g(\cdot) = \left(\frac{f_{j+1}(\cdot)}{(\ln(f_1(\cdot))f_1(\cdot))^2} - \frac{\Psi_j(f(\cdot))}{f_1(\cdot)}\right)g_1(\cdot)$$
$$- \frac{1}{jf_1(\cdot)} \sum_{i=1}^{j-1}(j-i)f_{2j-i+1}(\cdot)g_{i+1}(\cdot)$$
$$- \frac{1}{\ln(f_1(\cdot))f_1(\cdot)}g_{j+1}(\cdot) - \frac{1}{jf_1(\cdot)}\sum_{i=1}^{j-1}if_{j-i+1}(\cdot)g_{i+j+1}(\cdot).$$

We now proceed by induction. By Theorem 4.2,

$$(\hat{e}_{0,n}(\cdot), \ldots, \hat{e}_{m,n}(\cdot)) \Rightarrow (\hat{e}_0(\cdot), \ldots, \hat{e}_m(\cdot))$$

in $D_{\sigma,\phi}^{m+1}$. First, we deduce by the $\delta$-method (see Theorem 3.9.4 in [41]) that

$$\sqrt{k_n}(\chi_n^{(\cdot)}(1) - \pi(1)) \Rightarrow \Psi_1'(p^{(\cdot)}(0), p^{(\cdot)}(1)) \cdot (\hat{e}_0(\cdot), \hat{e}_1(\cdot)) = w_1(\cdot)$$

in $D_{\sigma,\phi}^1$. Then

$$\hat{d}_{1,n}(\cdot) = \max(-\sqrt{k_n}\pi(1), \min(\sqrt{k_n}(\chi_n^{(\cdot)}(1) - \pi(1)), \sqrt{k_n}(1 - \pi(1)))) \Rightarrow \hat{d}_1(\cdot)$$

in $D_{\sigma,\phi}^1$ and

$$(\hat{e}_{0,n}(\cdot), \hat{e}_{1,n}(\cdot), \hat{d}_{1,n}(\cdot)) \Rightarrow (\hat{e}_0(\cdot), \hat{e}_1(\cdot), \hat{d}_1(\cdot))$$

in $D_{\sigma,\phi}^3$. Assume that we have already shown that

$$(\hat{e}_{0,n}(\cdot), \ldots, \hat{e}_{j,n}(\cdot), \hat{d}_1(\cdot), \ldots, \hat{d}_{j-1}(\cdot)) \Rightarrow (\hat{e}_0(\cdot), \ldots, \hat{e}_j(\cdot), \hat{d}_1(\cdot), \ldots, \hat{d}_{j-1}(\cdot)).$$

The $\delta$-method yields

$$\sqrt{k_n}(\chi_n^{(\cdot)}(j) - \pi(j)) \Rightarrow \Psi_j'(p^{(\cdot)}(0), \ldots, p^{(\cdot)}(j), \pi^{(\cdot)}(1), \ldots, \pi^{(\cdot)}(j-1))$$
$$\times (\hat{e}_0(\cdot), \ldots, \hat{e}_j(\cdot), \hat{d}_1(\cdot), \ldots, \hat{d}_{j-1}(\cdot))$$

in $D_{\sigma,\phi}^1$, and a straightforward computation shows that the limit is equal to $w_j(\cdot)$. Let us recall that

$$\hat{d}_{j,n}(\cdot) = \max\bigg(-\sqrt{k_n}\pi(j),$$
$$\min\bigg(\sqrt{k_n}(\chi_n^{(\cdot)}(j) - \pi(j)), \sqrt{k_n}\bigg(1 - \sum_{i=1}^{j}\pi(i)\bigg) - \hat{\psi}_{j,n}(\cdot)\bigg)\bigg),$$

where $\hat{\psi}_{j,n}(\cdot) = \sum_{i=1}^{j-1} \hat{d}_{i,n}(\cdot)$. It follows that $\hat{d}_{j,n}(\cdot) \Rightarrow \hat{d}_j(\cdot)$ in $D_{\sigma,\phi}^1$, and

$$(\hat{e}_{0,n}(\cdot), \ldots, \hat{e}_{j+1,n}(\cdot), \hat{d}_1(\cdot), \ldots, \hat{d}_j(\cdot)) \Rightarrow (\hat{e}_0(\cdot), \ldots, \hat{e}_{j+1}(\cdot), \hat{d}_1(\cdot), \ldots, \hat{d}_j(\cdot))$$

in $D_{\sigma,\phi}^{2(j+1)}$. The induction is established and

$$(\hat{d}_{1,n}(\cdot), \ldots, \hat{d}_{m,n}(\cdot)) \Rightarrow (\hat{d}_1(\cdot), \ldots, \hat{d}_m(\cdot))$$

in $D_{\sigma,\phi}^m$. By the CMT, we deduce that

$$(\bar{d}_{1,n}, \ldots, \bar{d}_{m,n}) \xrightarrow{d} \bigg(\frac{1}{\phi - \sigma}\int_\sigma^\phi \hat{d}_1(\tau)\,d\tau, \ldots, \frac{1}{\phi - \sigma}\int_\sigma^\phi \hat{d}_m(\tau)\,d\tau\bigg). \quad \square$$

PROOF OF COROLLARY 4.2. The assertions follow from the $\delta$-method and the CMT. $\square$

**Acknowledgments.** We would like to thank the referees and the Associate Editor for their comments which have helped to improve several aspects of the paper and for drawing our attention to the references [5] and [6].

CONSERVATOIRE NATIONAL DES ARTS ET METIERS
CHAIRE DE MODELISATION STATISTIQUE
CASE 445
292 RUE SAINT MARTIN
75003 PARIS
FRANCE
E-MAIL: chrobert@ensae.fr